\documentclass[prb,aps,showpacs,amsmath]{revtex4}
\usepackage{graphicx}
\usepackage{bm,latexsym}
\usepackage{amstext,amssymb}
\usepackage{epsfig}
\newcommand{\mitwert}[1]{\left< #1 \right>}

\newcommand{\bea}{\begin{eqnarray}}
\newcommand{\eea}{\end{eqnarray}}
\newcommand{\be}{\begin{equation}}
\newcommand{\ee}{\end{equation}}
\newcommand{\bi}{\begin{itemize}}
\newcommand{\ei}{\end{itemize}}

\begin{document}
\title{Fundamental questions relating to ion conduction in disordered solids}
\author{Jeppe C. Dyre}
\affiliation{DNRF centre  ``Glass and Time,'' IMFUFA, Department of Sciences, Roskilde University, Postbox 260, DK-4000 Roskilde, Denmark}
\author{Philipp Maass}
\affiliation{Institut f{\"u}r Physik, Technische Universit{\"a}t Ilmenau, D-98684 Ilmenau, Germany;\\
Fachberecih Physik, Universit\"at Osnabr\"uck, Barbarastra{\ss}e 7,D-49069 Osnabr\"uck, Germany}
\author{Bernhard Roling}
\affiliation{Fachbereich Chemie, Philipps-Universit{\"a}t Marburg, Hans-Meerwein Str., D-35032 Marburg, Germany}
\author{David L. Sidebottom}
\affiliation{Department of Physics, Creighton University, Omaha, NE-68178, USA}
\date{\today}

\begin{abstract}
A number of basic scientific questions relating to ion conduction in homogeneously disordered solids are discussed. The questions deal with how to define the mobile ion density, what can be learned from electrode effects, what is the ion transport mechanism, the role of dimensionality, and what are the origins of the mixed-alkali effect, of time-temperature superposition, and of the nearly-constant loss. Answers are suggested to some of these questions, but the main purpose of the paper is to draw attention to the fact that this field of research still presents several fundamental challenges.
\end{abstract}


\maketitle

\section{Introduction}

Ion conduction in glasses, polymers, nanocomposites, highly defective crystals, and other disordered solids plays an increasingly important role in technology. Considerable progress has been made recently, for instance with solid-oxide fuel cells, electrochemical sensors, thin-film solid electrolytes in batteries and supercapacitors, electrochromic windows, oxygen-separation membranes, functional polymers, etc.\cite{kna00,dub03,kha04,kna04,vin06,hui07,nik07,gro08,fun08} The applied perspective is an important catalyst for work in this field. In this paper, however, the focus is on basic scientific questions. This is relevant because ion transport in disordered materials remains poorly understood. There is no simple, broadly accepted model; it is not even clear whether any generally applicable, simple model exists. Given the intense current interest in the field -- with hundreds of papers published each year -- it is striking that there is no general consensus on several fundamental questions.\cite{bun98} This is in marked contrast to other instances of electrical conduction in condensed matter where a much better understanding has been achieved, e.g., for electronic conduction in metals, semiconductors, and superconductors, as well as for ion conduction by defects in crystals.

This paper summarizes and discusses basic scientific questions relating to ion conduction in (mainly) homogeneously disordered solids \cite{owe63,tom77,tul80,ing87,vin87,kre89,ang90,mar91,ang92,hei03,mai95}. The main motivation is not to suggest or provide answers, but to inspire to further research into the fundamentals of ion conduction in disordered solids. A question that is not addressed below, which has been a point of controversy particularly during the last decade, is how to best represent ac data, via the conductivity or the electric modulus,\cite{alm1983,dyr91,ell94,moy94} . By now this has been thoroughly discussed in the literature, and we refer the interested reader to the discussion in Refs. \onlinecite{nga00,sid01,hod05} that present and summarize the differing viewpoints.

\begin{figure}
\includegraphics[width=8cm]{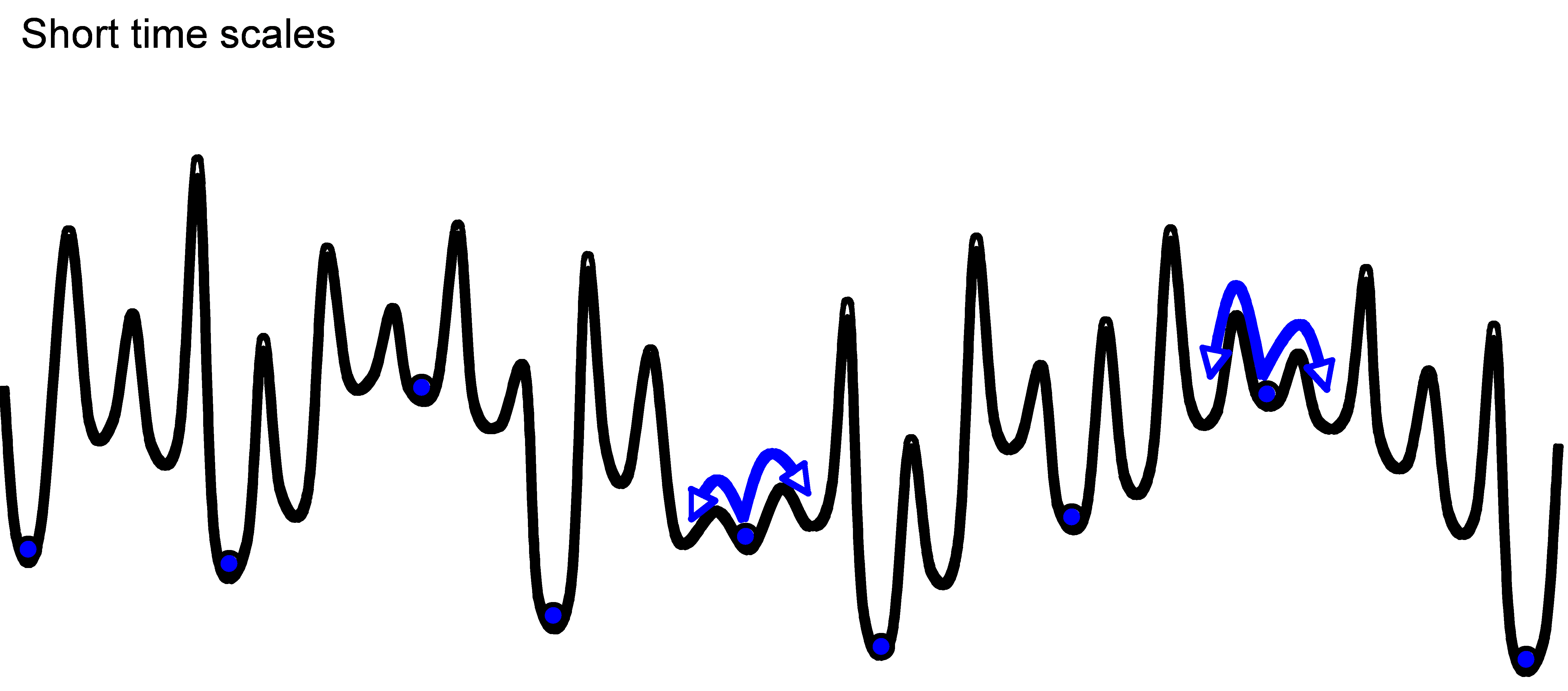}
\includegraphics[width=8cm]{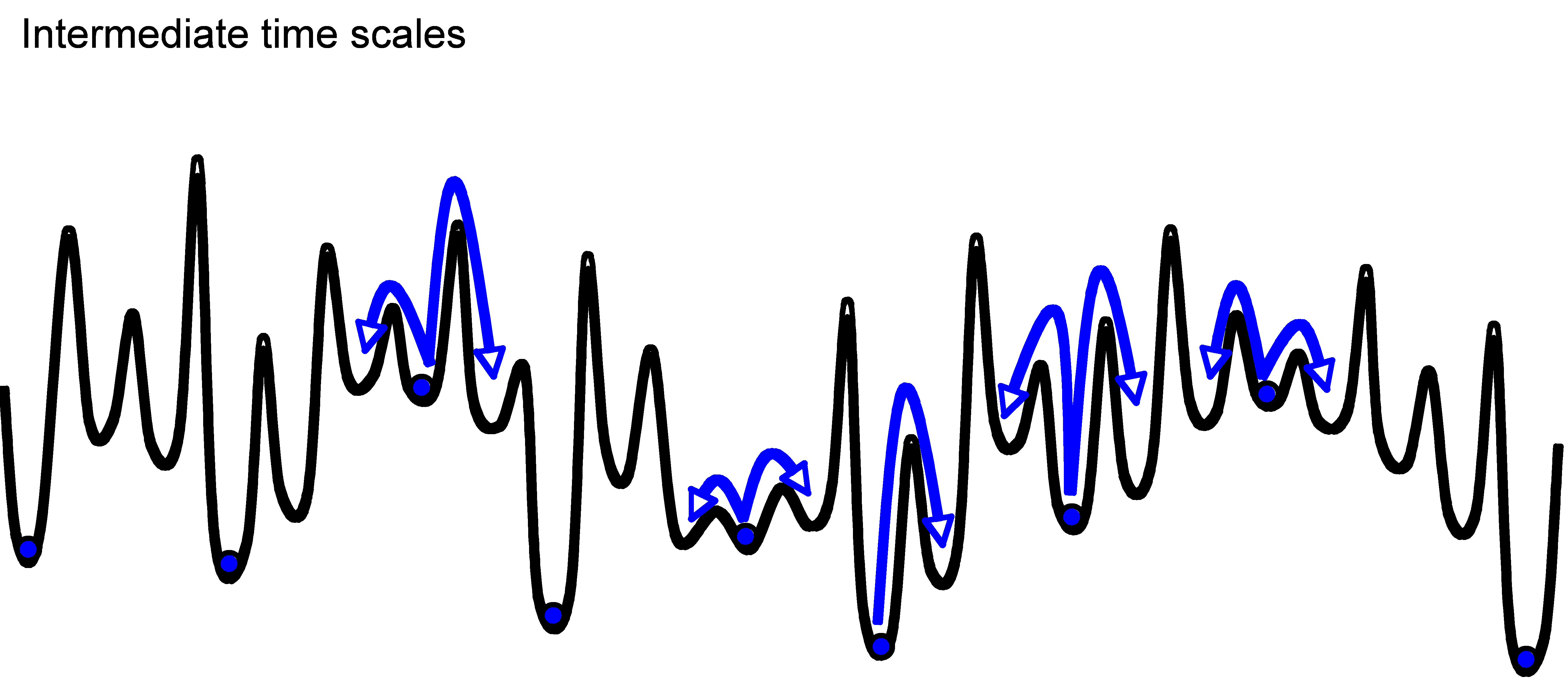}
\includegraphics[width=8cm]{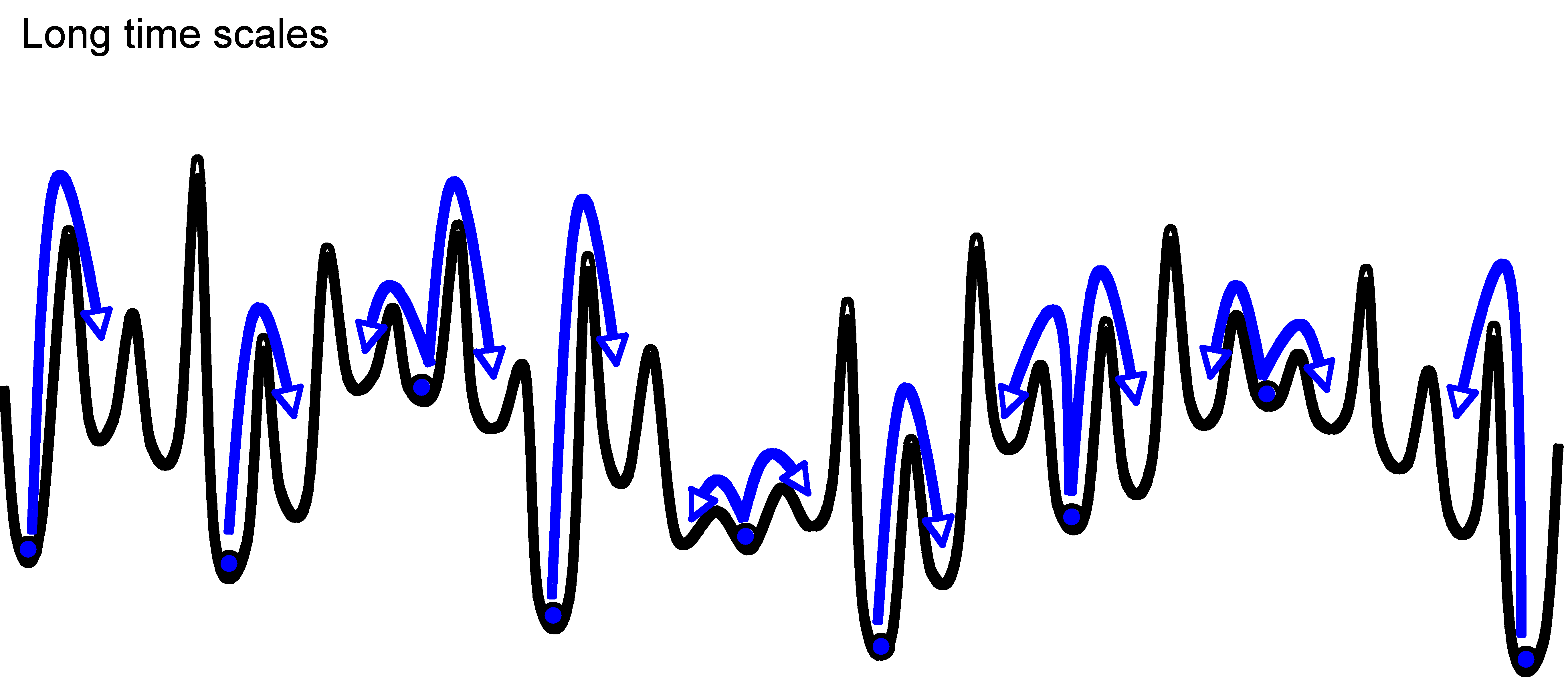}
\caption{Schematic figures illustrating ion jumps in a disordered landscape, here in one dimension. The arrows indicate attempted jumps. Most of these are unsuccessful and the ion ends back in the minimum it tried to leave: If the barrier is denoted by $\Delta E$, if $T$ is the temperature and $k_B$ is Boltzmann's constant, according to rate theory the probability of a successful jump is roughly $\exp(-\Delta E/k_BT)$. This implies that on short time scales only the smallest barriers are surmounted. As time passes, higher and higher barriers are surmounted, and eventually the highest barriers are overcome. In more than one dimension the highest barrier to be overcome for dc conduction is determined by percolation theory; there are even higher barriers, but these are irrelevant because the ions go around them.}
\end{figure}

\section{How to define mobile ion density?}

Ion motion in disordered solids is fundamentally different from electronic conduction in crystalline solids. Ions are much heavier than electrons so their motion is far less governed by quantum mechanical descriptions. Below typical vibrational frequencies ($\lesssim100$~GHz) ion motion can be described by activated hopping between (usually) charge compensating sites. Moving ions carry charge, of course, and thus produce an electrical response which can be detected by a variety of experimental techniques. Unlike crystals, the potential energy landscape experienced by an ion in a glass or otherwise disordered solid is irregular and contains a distribution of effective depths and barrier heights, as sketched in Fig. 1. The effective energies result from differing binding energies at residence sites and differing saddle point energies between residence sites, and they are influenced by interactions between the ions. With increasing time scale, the ions can explore larger parts of space by overcoming higher energy barriers.

Following standard arguments, suppose ions with charge $q$ are subjected to an electric field $E$. The field exerts the force $qE$ on each ion, resulting in an average drift velocity $v$ in the field direction. The ion mobility $\mu$ is defined by $\mu=v/E$.  If the number of mobile ions per unit volume is $n_{\rm mob}$, the current density $J$ is given by $J=qn_{\rm mob}v$. Thus we obtain the following expression for the dc conductivity defined by $\sigma_{dc}\equiv J/E$: 

\begin{equation}\label{mobility} \sigma_{dc}
\,=\,q\,n_{\rm mob}\,\mu\,\,.
\end{equation}
This equation expresses the simple fact that the conductivity is proportional to the ion charge, to the number of mobile ions, and to how easily an ion is moved through the solid. As such, Eq. (\ref{mobility}) is an excellent starting point for discussing how the conductivity depends on factors like temperature and chemical composition. Or is it? We shall now argue that the above conventional splitting of the conductivity into a product of mobility and mobile ion density involves non-trivial assumptions.

Except at very high temperatures ion motion in solids proceeds via jumps between different ion sites. Most of the time an ion vibrates in a potential-energy minimum defined by the surrounding matrix. This motion does not contribute to the frequency-dependent conductivity except at frequencies above the GHz range; only ion jumps between different minima matter. The mobility reflects the long-time average ion displacement after many jumps. The fact that ions spend most of their time vibrating in potential energy minima, however, makes the definition of mobile ion density less obvious: How to define the number of mobile ions when all ions are immobile most of the time?

Intuitively, Eq. (\ref{mobility}) still makes sense. Imagine a situation where some ions are very tightly bound (``trapped'') while others are quite mobile. In this situation one would obviously say that the density of mobile ions is lower than the total ion density. The problem, however, is that the tightly bound ions sooner or later become mobile, and the mobile ions sooner or later become trapped: By ergodicity, in the long run all ions of a given type must contribute equally to the conductivity. Thus on long time scales the ``mobile'' ion density must be the total ion concentration. This ``long run'' may be years or more, and ions trapped for so long are for all practical purposes immobile. Nevertheless, unless there are infinite barriers in the solid, which is unphysical, in the very long run all ions are equivalent.

The question how many ions contribute to the conductivity makes good sense, however, if one specifies a time scale. Thus for a given time $\tau$ it is perfectly well-defined to ask: On average, how many ions move beyond pure vibration within a time window of length $\tau$? If the average concentration of ions moving over time $\tau$ is denoted by $n_{\rm mob}(\tau)$ and $n$ is the total ion concentration, ergodicity is expressed by the relation

\begin{equation}\label{n_tau_def}
n_{\rm mob}(\tau\rightarrow\infty)\,=\,n\,.
\end{equation}

An obvious question is how to determine $n_{\rm mob}(\tau)$ experimentally. A popular method of determining the ``mobile ion density'' -- without explicit reference to time scale -- is by application of the Almond-West (AW) formalism\cite{alm83a,alm83b,hai94} that takes advantage of the frequency dependence of the conductivity. We proceed to discuss this approach. First note that in ion conductors with structural disorder, the short-time ion dynamics is characterized by back-and-forth motion over limited ranges, ``subdiffusive'' dynamics, whereas the long-time dynamics is characterized by random walks resulting in long-range ion transport, ``diffusive'' dynamics (Fig. 1).\cite{dyr88,maa91,fun93,maa95,dyr00} The back-and-forth motion leads to dispersive conductivity at high frequencies, while the long-range transport leads to the low-frequency plateau marking the dc conductivity (Fig. 2). There is experimental evidence that in materials with high ion concentration, at any given time only part of the ions are actively involved in back-and-forth motion.\cite{rol01,mur04}.

\begin{figure}
\includegraphics[width=8cm]{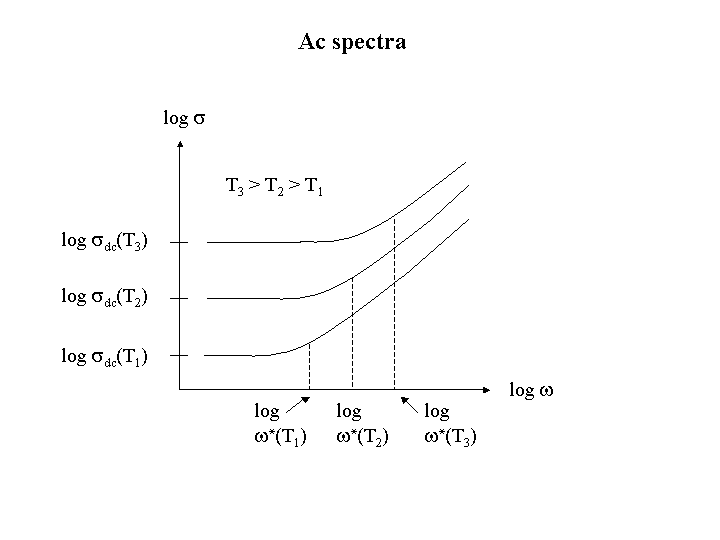}
\includegraphics[width=8cm]{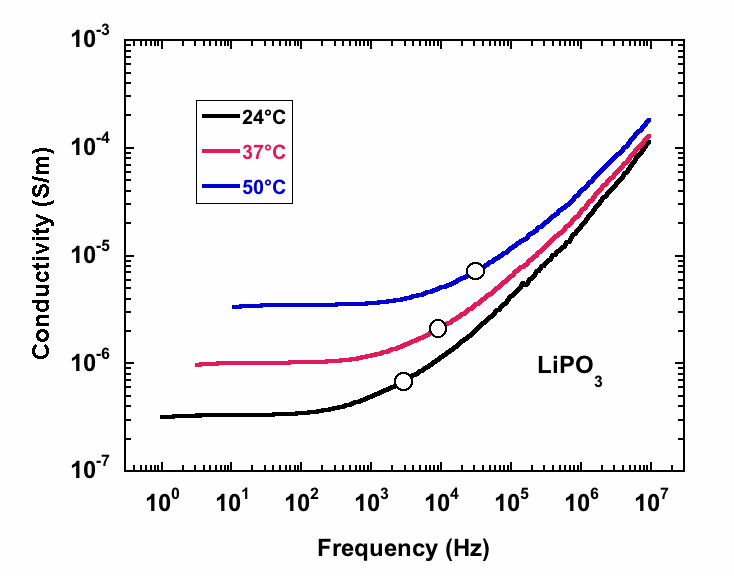}
\caption{(a) Schematic figure showing the real part of the ac conductivity as a function of frequency at three different temperatures. As temperature is lowered, the dc conductivity decreases rapidly. At the same time the frequency marking onset of ac conductivity also increases (actually in proportion to the dc conductivity, compare to the BNN relation Eq. (\ref{eq:bnn}) discussed below).
(b) The real part of the conductivity at three different temperatures for a Lithium-Phosphate glass.}
\end{figure}

A widely applied description of conductivity spectra in the low-frequency regime (i.e., below 100~MHz) is a Jonscher type power law, 

\begin{equation}\label{jonscher}
\sigma'(\nu)=\sigma_{\rm dc}\left[1+\left(\frac{\nu}{\nu^\ast}
\right)^n\;\right]
\end{equation}
where we have written the equation in a form such that the crossover frequency marking the onset of ac conduction, $\nu^\ast$, is given by $\sigma'(\nu^\ast)=2\sigma_{\rm dc}$. Equation~(3) is sometimes referred to as the Almond and West (AW) formula, although Almond and West did not consider Jonscher's ``universal dielectric response'' of disordered systems, but introduced their formula to describe defective crystals with an activated number of charge carriers. Nevertheless, when applying Eq.~(3) to strongly disordered systems, as, e.g.\ ionic glasses, many authors in the literature follow the physical interpretation suggested by Almond and West and identify the crossover frequency with a ``hopping rate''. Thus combining this ansatz with the Nernst-Einstein relation gives 

\begin{equation}
n_{\rm AW}=\frac{6k_{\rm B}T}{q^2a^2}
\frac{\sigma_{\rm dc}}{\nu^\ast}
\label{eq:nmob-aw}
\end{equation}
as an equation to determine the number density of ``mobile ions'', $n_{\rm AW}$ (after typically assuming jump lengths $a=2-3$~\AA).

However, if one accepts that Eq. (\ref{jonscher}) provides a good fit to spectra in the low-frequency regime -- it generally fails at frequencies above 100~MHz --  the estimate of an effective number density of ``mobile ions'' based on Eq. (\ref{eq:nmob-aw}) is questionable. Application of the fluctuation-dissipation theorem  implies the following expression, where $t^\ast\equiv 1/\nu^\ast$ and $H$ and $\gamma$ are numbers that are roughly of order unity ($H$ is an in principle time-scale-dependent Haven ratio \cite{hav65,ter75,isa99} reflecting ion-ion correlations and $\gamma\sim 2$ is a numerical factor reflecting the conductivity spectrum at onset of ac conduction, see Appendix A):

\be
\label{eq_sigdc1}
\sigma_{dc} \,=\, \frac{n\,q^2}{6\,k_B T}  \frac{\mitwert{\Delta {\bf r}^2(t^{\ast})}}{\gamma\,H}\,\nu^{\ast}\,.
\ee	
Combining Eqs. (\ref{eq:nmob-aw}) and (\ref{eq_sigdc1}) yields

\be
\label{eq_NV}
\frac{n_{ \rm AW}}{n} \,=\, \frac{1}{\gamma\,H}\frac{\mitwert{\Delta {\bf r}^2(t^{\ast})}}{a^2}\,.
\ee
If the mean-square displacement obeys $\mitwert{\Delta {\bf r}^2(t^{\ast})} \simeq a^2$ one has $n_{ \rm AW}\simeq n_{\rm mob}(\tau=1/\nu^\ast)$, but unfortunately the quantity $\mitwert{\Delta {\bf r}^2(t^{\ast})}$ does not generally have this approximate value. In a simple models where all ions have similar jump rate, $\mitwert{\Delta {\bf r}^2(t^{\ast})}$ is indeed roughly $a^2$ times the fraction of ions that have jumped within time $t^\ast$. It is not possible to model the universally observed strong frequency dispersion of the conductivity without assuming a wide spread of jump rates, however, and in such models like the random barrier model (RBM) considered below $\mitwert{\Delta {\bf r}^2(t^{\ast})}$ is much larger than $a^2$. Generally, $\mitwert{\Delta {\bf r}^2(t^{\ast})}/a^2$ gives an approximate {\it upper limit} for the fraction of ions that have moved in the time window $t^\ast$. Ignoring the less significant factor $\gamma H$, this implies that $n_{\rm mob}(t^{\ast}) <  n_{ \rm AW}$. To summarize, only in models where all ions have similar jump rates does $n_{ \rm AW}$ give an estimate of how many ions on average jump over the time interval of length $t^{\ast}$.

An alternative suggestion for obtaining information about the ``number of mobile ions'' is based on analyzing the electrode polarization regime of conductivity 
spectra for ion conductors placed between blocking electrodes.\cite{sch94,tom98,kle06,mar08} However, theoretical analyses of the spectra are often based on Debye-H{\"u}ckel-type approaches,\cite{sch94,tom98,kle06,mar08} the applicability of which is far from obvious at high ion density. Thus while it is a potentially useful idea, more theoretical work is needed before observations of electrode effects may lead to safe conclusions regarding the number of mobile ions (see the next section that outlines the a simple approximate description); one still needs to specify the time scale that the number of mobile ions refers to. -- Solid-state NMR methods such as motional narrowing experiments\cite{riv04,ber05,mus06} and the analysis of multi-time correlation functions of the Larmor frequency,\cite{boh03,boh07} provide information about the number of ions moving on the time scale that these methods monitor (milliseconds to seconds). 

The question ``what is the density of mobile ions?'' is thus well defined only when it refers to a particular time scale. This is because according to standard ergodicity arguments, if the time scale is taken to infinity, all ions contribute equally and the density of mobile ions is the total ion density $n$. A natural choice of time scale is that characterizing the onset of ac conduction, the $t^\ast$ of the above equations. Choosing this time scale leads to a classification of ion conductors into two classes: Those for which $n_{\rm mob}(t^\ast)$ is comparable to the total ion density $n$: $n_{\rm mob}(t^\ast)\simeq n$ (``strong electrolyte case'' \cite{tul80}), and those for which $n_{\rm mob}(t^\ast)\ll n$ (``weak electrolyte case'' \cite{rav77,ing80}). The latter class includes solids where ion conduction proceeds by the vacancy mechanism (Sec. VII).

\section{What can be learnt from electrode polarization?}  

As is well known, the ac conductivity $\sigma(\omega)$ is a complex function. Thus associated with the real part there is also an imaginary component; the latter determines the real part of the frequency-dependent permittivity. For the study of ion conduction in disordered solids the use of blocking or partially blocking metal electrodes is convenient. In this case, the high-frequency parts of ac conductivity and permittivity spectra are governed by ion movements in the bulk of the solid electrolyte, while the low-frequency part is governed by so-called ``electrode polarization'' effects, as shown in Fig. 3. Since the ions are blocked by the metal electrode, there is accumulation or depletion of ions near the electrodes, leading to the formation of space-charge layers. The voltage drops rapidly in these layers, which implies a huge electrical polarization of the material and a near-absence of electric field in the bulk sample at low frequencies. The build-up of electrical polarization and the drop of the electric field in the bulk are reflected in an increase of the ac permittivity and a decrease of the ac conductivity with decreasing frequency.\cite{isa76} For completely blocking electrodes $\sigma(0)=0$, of course. -- Whenever both ions and electrons conduct, a number of electrochemical techniques exist for evaluating transference numbers of ions and electrons, including galvanic cells, polarization, and permeation techniques.\cite{kna04,hey77,rie91,rie97}

\begin{figure}
\includegraphics[width=8cm]{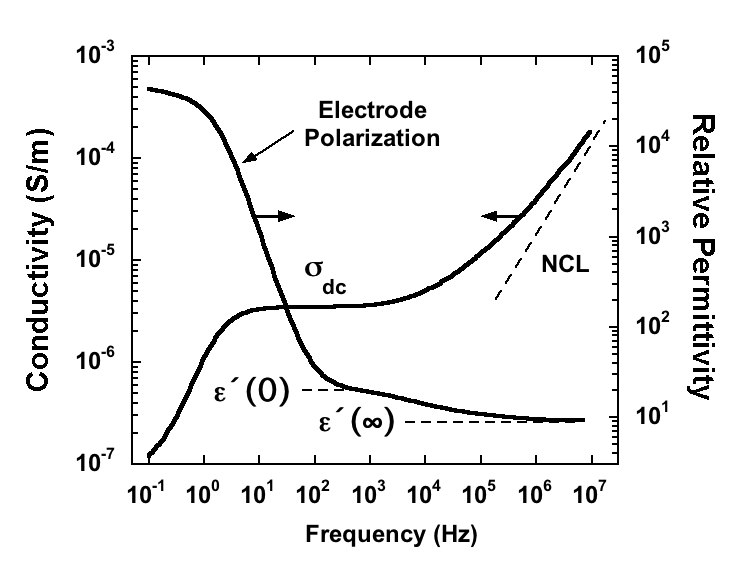}
\includegraphics[width=8cm]{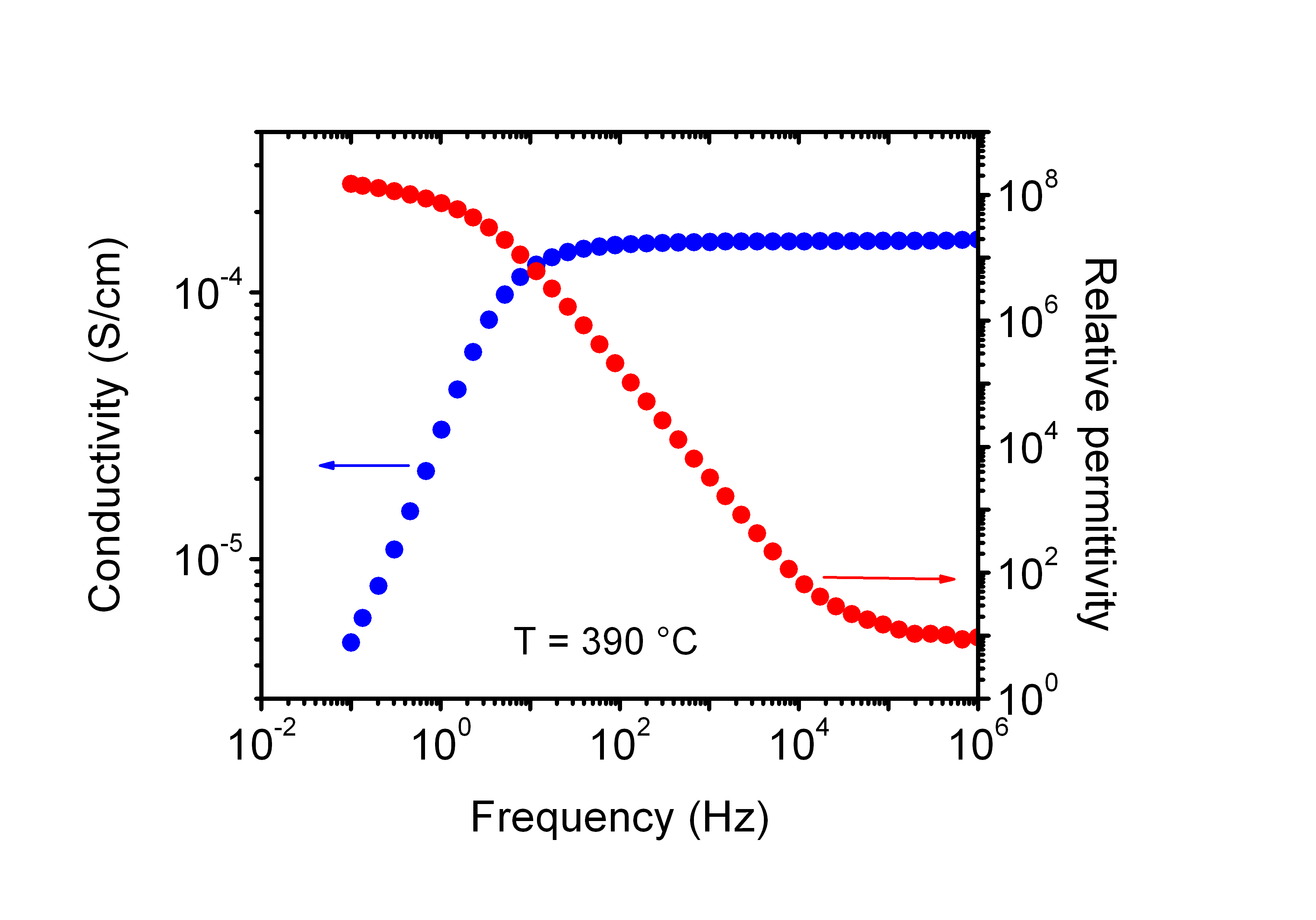}
\caption{(a). At high frequencies the nearly constant loss (NCL) regime appears where the conductivity becomes almost proportional to frequency, sometimes even closer to a straightforward proportionality than shown here (data for a Lithium-Phosphate glass).
(b) The electrode polarization effects on the real part of the conductivity and the real part of the dielectric constant at high temperature for a Na-Ca-phosphosilicate glass.}
\end{figure}

Systematic experimental and theoretical studies of electrode polarization effects in electrolytes began in the 1950s in works was carried out by Macdonald \cite{mac53}, Friauf \cite{fri54}, Ilschner \cite{ils58}, Beaumont \cite{bea67}, and others.  Their approaches were based on differential equations for the motion (diffusion and drift) of charge carriers under the influence of chemical and electrical potential gradients. These equations were combined with the Poisson equation and linearized with respect to the electric field. Thereby, expressions for the ac conductivity and permittivity at low electric field strengths were derived. These are mean-field approaches in the sense that a mobile charge carrier interacts with the average field produced by the electrode and the other mobile carriers \cite{baz04}. 
 
\begin{figure}
\includegraphics[width=12cm]{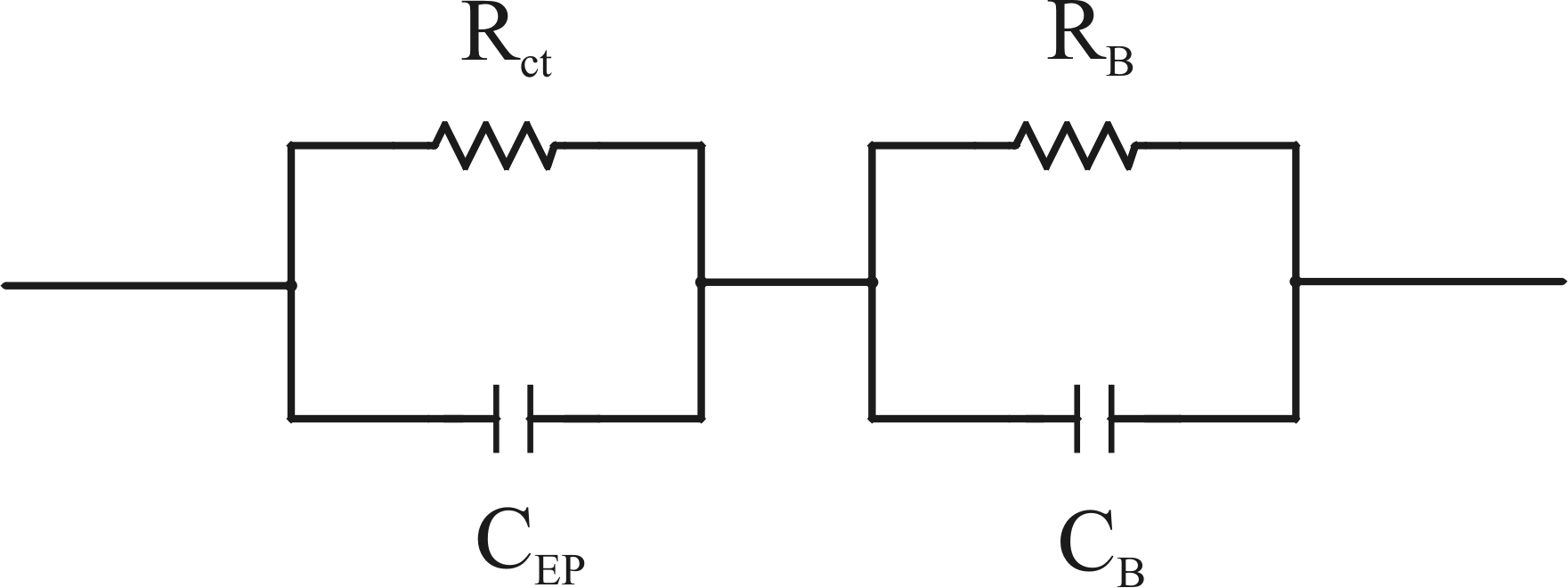}
\caption{Simplified electrical equivalent circuit describing the low-field ac conductivity and permittivity spectra of solid electrolytes between blocking ($R_{ct}=\infty$) or partially blocking electrodes. The right element describes the bulk properties, the left describes the space charge layer capacitance and charge transfer resistance (the frequency dispersion of the bulk conductivity may be taken into account by replacing the resistor $R_B$ by a frequency-dependent impedance).}
\end{figure}

When charge carrier formation and recombination can be neglected and the sample thickness $L$  is much larger than the space-charge layer thickness, the theoretical expressions can be approximately mapped onto the simple electrical equivalent circuit shown in Fig. 4 if the frequency dependence of the bulk conductivity is ignored. Ion transport in the bulk is described by the $R_BC_B$ element. The space charge layers are described by a capacitance $C_{EP}$ , and, in case of a discharge of the mobile ions at the electrode, a (generally large) parallel charge transfer resistance $R_{ct}$ . The $R_{ct}C_{EP}$ element acts in series with the $R_BC_B$ element. In the cases $C_{EP}\gg C_B$ and $R_{ct}\gg R_B$ that usually apply, the equivalent circuit leads to the following expressions for the frequency-dependent conductance $G(\omega)$ (real part of the admittance) and capacitance $C(\omega)$ (imaginary part of the admittance):

\begin{equation}\label{elec1}
G(\omega)\,\equiv\,
\sigma'(\omega)\frac{A}{L}\,=\,
\frac{(1/R_{ct}+R_B/R_{ct}^2)+\omega^2R_BC_{EP}^2}
{(1+R_B/R_{ct})^2+(\omega R_BC_{EP})^2}
\end{equation}
and

\begin{equation}\label{elec2}
C(\omega)\,\equiv\,
\epsilon_0 \epsilon'(\omega)\frac{A}{L}\,=\,
C_{EP}\frac{1+\omega^2 R_B^2 C_B C_{EP}}{(1+R_B/R_{ct})^2+(\omega R_BC_{EP})^2}
\end{equation}
with $C_{EP}/C_B=L/(2L_D)$ and $A$ denoting the sample area, where the Debye length $L_D$ is defined by

\begin{equation}\label{elec3}
L_D^2\,\equiv\,
\frac{\epsilon_0\epsilon_{\rm bulk} k_BT}{\tilde{n}_{\rm mob} e^2}\,.
\end{equation}
From these expressions a number density of mobile ions $\tilde n_{\rm mob}$ can be calculated that is the density of mobile ions referring to the time scale for build up of electrode polarization, $\tilde n_{\rm mob}=n_{\rm mob}(\tau_{\rm ep})$ where $\tau_{\rm ep}=R_{B}C_{EP}$.

In the absence of ion discharge, i.e., when $R_{ct}\rightarrow\infty$ , the equivalent circuit reduces to an RC element in series with a capacitor. The existence of a finite charge transfer resistance leads to the occurrence of a conductance plateau at low frequencies with plateau value $G_s$ given by

\begin{equation}\label{elec4}
G_s\,=\,
\frac{1}{R_B+R_{ct}}\,.
\end{equation}
In addition, the static capacitance $C_s$  becomes slightly smaller than $C_{EP}$:

\begin{equation}\label{elec5}
C_s\,=\,
C_{EP}\left(\frac{R_{ct}}{R_{ct}+R_B}\right)^2\,.
\end{equation}
                                                     
This mean-field approach should apply to materials with low $\tilde{n}_{\rm mob}$, such as ionic defect crystals and diluted electrolyte solutions. Its applicability to disordered solids with high ion density is far from obvious. Nevertheless, quite a number of ac spectra of ion conducting glasses and polymers were traditionally analyzed and interpreted utilizing the above equations. Thereby, number densities of mobile ions were calculated and compared to the total ion content of the samples. For instance, Sch{\"u}tt and Gerdes concluded that in alkali silicate and borosilicate glasses only between 1 ppm and 100 ppm of the alkali ions are mobile \cite{sch92}. Similar results were obtained by Tomozawa on silica glass with impurity ions \cite{tom98} and by Pitarch et al. from voltage-dependent measurements of on a sodium aluminosilicate glass \cite{pit03}. Klein et al. carried out measurements on ionomers containing alkali ions and found ratios of mobile alkali ions to the total alkali ion content ranging from about 10 ppm to 500 ppm \cite{kle06}. 

For a critical discussion of such experimental results and their interpretation, it is important to consider limitations of both experiment and theory. Regarding the experimental situation there are in particular two important points: (i) The roughness of the electrode/solid electrolyte interface is usually not taken into account. Especially in a frequency range where the length scale of the potential drop at the electrodes is comparable to the roughness of the interface, the roughness must have a considerable influence on the ac conductivity and permittivity. (ii) The surface-near regions of ion conductors often exhibit a chemical composition that is significantly different from the bulk. For instance, in ionic glasses, surface corrosion is initiated by an alkali-proton exchange. Such deviations from the bulk composition should have a strong influence on the ac spectra when the potential drop occurs very close to the surface, i.e., at high capacitance values close to the static capacitance plateau and in the static capacitance plateau regime.

From a theoretical point of view, serious limitations of the applicability of mean-field approaches to disordered solids derive from: (i) The interactions between the ions are not taken into account; (ii) surface space charges in disordered solids may exist even without the application of an external electric field, due to ion exchange processes at the surface or due to an interaction of mobile ions with the metal electrode. Thus more sophisticated theories should take into account the possibility of an open-circuit potential difference between electrodes and solid electrolyte.

In summary, considerable efforts in both experiment and theory is required in order to carry out measurements on well-defined electrode/electrolyte interfaces and to obtain a better theoretical understanding of what kind of information may be derived from electrode polarization effects. It is clearly worthwhile to pursue this direction of research, and it would also be worthwhile to look into what can be learned from electrode effects in the strong-field case where the electrode polarization becomes nonlinear.

\section{What causes the mixed-alkali effect?}

A prominent phenomenon occurring in ion-conducting glasses is the mixed-alkali effect (for reviews, see \cite{hug68,isa68,day76,ing94}). This effect is the increase of the mobility activation energy of one type of ion when it is gradually replaced by a second type of mobile ion. This leads to changes of the tracer diffusion coefficients $D_{A,B}(x)$ over several orders of magnitude at low temperatures, and to a minimum in the dc-conductivity $\sigma(x)\sim (1-x)D_A(x)+xD_B(x)$, when the mixing ratio $x$ of two mobile ions $A$ and $B$ is varied (for recent systematic experimental studies, see e.g. Refs. \onlinecite{vos04a,vos04b}).

Much progress has been made over the last two decades for explaining the mixed-alkali effect 
\cite{maa92,bun94,hun94,gre95,hab95,hun95,hab96,swe98,tom98b,bar99,maa99,sch99,kir00,swe01,swe03,bun04,hab04b,imr06,hab07b,zie08}. Compelling evidence now exists that its origin is of structural character, associated with a mismatch effect \cite{maa92} where sites in the glassy network favorable for one type of mobile ion are unfavorable for the other type of mobile ion. This evidence comes from EXAFS \cite{gre95,gre91,hou93}, NMR \cite{gee97} and infrared spectroscopy \cite{kam93,kam96,kam98}, x-ray and neutron scattering experiments in combination with reverse Monte-Carlo modeling \cite{swe01} and bond valence sum analyses \cite{swe03}, molecular orbital calculations \cite{uch92,uch99}, molecular dynamics simulations \cite{hab95,bal93,lam03,lam05} and theoretical work based on microscopic and semi-microscopic approaches \cite{maa92,hun94,maa99,bar99,kir00}. In hopping systems, the mismatch effect can be modeled by site energies that are different for different types of mobile ions, i.e., a low-energy site for one type is a high-energy site for the other type. 

Recently it was possible also to explain the peculiar behavior of the internal friction in mixed-alkali glasses \cite{pei05,maa06a}. When a mixed-alkali glass fibre is twisted at a certain frequency, two mechanical loss peaks can be identified well below the glass-transition temperature: the single-alkali peak that with beginning replacement becomes smaller and moves to higher temperatures, and the mixed-alkali peak that at the same time becomes higher and moves to lower temperatures (for a review of experimental results, see \cite{zda79}). Based on general theoretical considerations it was shown that the mixed-alkali peak can be traced back to mutual exchanges of two types of mobile ions and the single-alkali peak to exchanges of the (majority) ion with vacancies. As a consequence, large mixed-alkali peaks are predicted for ion types with small mismatch where ion-ion exchange processes are more likely to occur. This agrees with experimental observations. Moreover, it could be shown that the occurrence of large mixed-alkali peaks at small mixing ratios can be understood if the fraction of empty sites is small. This gives independent evidence for the small fraction of empty sites found in theoretical arguments \cite{dyr03} as well as in molecular dynamics simulations \cite{lam03,lam05,hab04,vog04b} (section VII).

Despite this progress over the past years, there are still many issues awaiting experimental clarification and theoretical explanation. A point less addressed so far in the microscopic modelling is the behavior of the viscosity as reflected in a minimum of the glass-transition temperature upon mixing. This softening of the glass structure at intermediate mixing ratios may significantly influence ion transport properties. The mixed-alkali effect becomes weaker with total mobile ion content \cite{day76} in agreement with theoretical expectations.\cite{maa92} However, a systematic theoretical study of this feature has not yet been undertaken. Overall, there is still no consistent theoretical account of all main signatures of the mixed-alkali effect.

We finally note that a mixed-alkali effect also occurs in crystals with structure of $\beta$- and $\beta''$-alumina type, where the ion motion is confined to two-dimensional conduction planes.\cite{cha78,fos81,bru83} A quantitative theory has been developed for this based on the wealth of structural information available.\cite{mey96,mey98} The key point is that $A$ and $B$ ions have different preference to become part of mobile defects, and this preference is caused by a different interaction of the ions with the local environment. In this respect the origin of the mixed-alkali effect in crystals has similarities to that in
glasses. However, different from the host network in glasses, the host lattice in the crystalline systems is almost unaffected by the mixing of the two types of mobile ions.

\section{What is the origin of time-temperature superposition?}  

Different suggestions were made in the past to characterize the similar ac responses observed for different types of ion conductors in frequency regimes not exceeding $\sim$100~MHz.  The simplest description is the power-law frequency dependence proposed by Jonscher (Eq. (\ref{jonscher})).\cite{jon77,jon83} The power-law description is not accurate, however, because the exponent must generally increase somewhat with frequency in order to fit experiment properly, and also because the asymptotic low-frequency behavior is inconsistent with experiment that imply $\sigma'(\omega)-\sigma(0)\propto\omega$ for $\omega\rightarrow 0$ ($\omega=2\pi\nu$).\cite{dyr00,fun07} A more general approach is to consider the scaling associated with time-temperature superposition (TTS) for any particular ion conductor \cite{bow06,mur07,pap07}. The scaling ansatz reads

\begin{equation}
\sigma(\omega,T)=\sigma_{\mathrm dc}(T)\, f\left[\omega/\omega^{\ast}(T)\right]\,.
\label{eq:scaling}
\end{equation}
Here $f(u)$ is the so-called scaling function and $\omega^{\ast}$  the previously defined angular frequency marking onset of ac conduction. Any solid that obeys TTS is, equivalently, referred to as obeying scaling. As an example Fig. 5(b) illustrates how the spectra of Fig. 2(b) scale to a common so-called master curve.

\begin{figure}
\includegraphics[width=8cm]{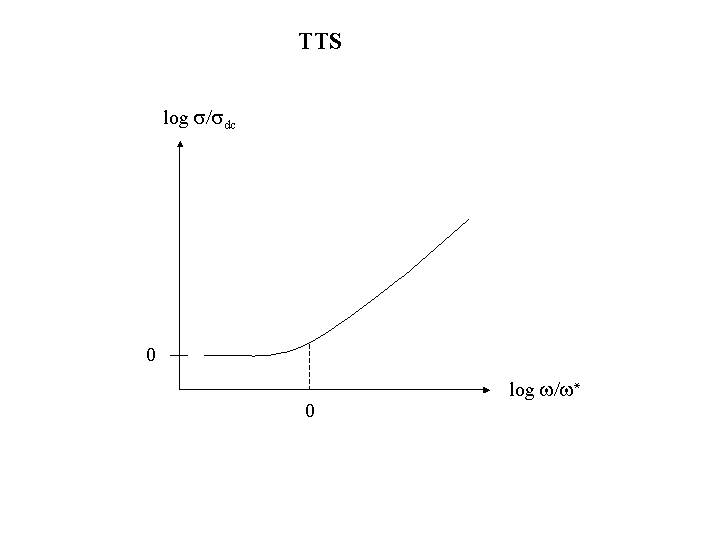}
\includegraphics[width=8cm]{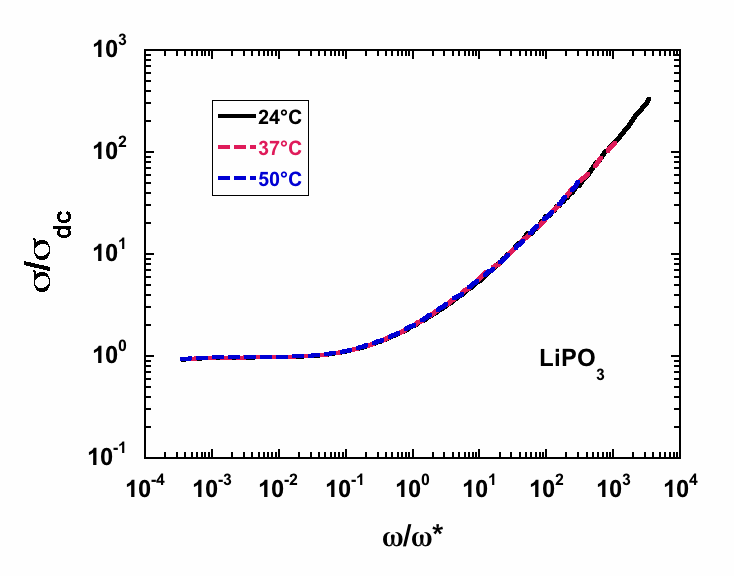}
\caption{Time-temperature superposition (TTS). (a) Sketch of how the three spectra of Fig. 2(a) collapse to a single master curve when suitably scaled. Whenever this is possible, the solid is said to obey TTS. Most disordered ion conductors, including all single-ion conducting glasses, obey TTS.
(b) TTS demonstrated for the Lithium-Phosphate glass ac data of Fig. 2(b).}
\end{figure}

Barton,\cite{bar66} Nakijama,\cite{nak72} and Namikawa\cite{nam75} long ago verified that for many ion- (and some electron-) conducting disordered solids 

\begin{equation}\label{eq:bnn} \omega^{\ast}=\frac{\sigma_{\mathrm dc}}{p\,\epsilon_0\,\Delta\epsilon},
\end{equation}
where $p$ is a constant of order of unity and $\Delta\epsilon$ is the dielectric strength, i.e., the difference between static and high-frequency dimensionless dielectric constants. Equation~(\ref{eq:bnn}) is known as the BNN relation.\cite{dyr86} By considering the low-frequency expansion of the conductivity a connection can be made between the scaling behavior Eq. (\ref{eq:scaling}) and the BNN relation.\cite{sch00} The argument assumes analyticity of the scaling function $f(u)$ for small $u$, which is in fact necessary in order to have a well-defined dielectric strength.\cite{die02} One has $\sigma(\omega)/\sigma_{\mathrm dc}\sim 1+iK\omega/\omega^\ast$ for $\omega\to 0$ with the constant $K$ being real. Accordingly, one obtains $\epsilon(\omega)-\epsilon_\infty\equiv\sigma(\omega)/(i\epsilon_0 \omega)\sim \sigma_{\mathrm dc}/(i\epsilon_0\omega)+K\sigma_{\mathrm dc}/ \epsilon_0\omega^{\ast}$ for $\omega\to0$ which implies $\Delta\epsilon=K\sigma_{\mathrm dc}/\epsilon_0\omega^{\ast}$. Thus TTS and analyticity imply the BNN relation -- but do not mathematically garantee that $p\sim 1$.

To the best of our knowledge, TTS applies for all single-ion conducting glasses and crystals with structural disorder. This remarkable fact suggests that disorder is intimately linked to TTS. In crystals with structural disorder, such as RbAg$_4$I$_5$ and $\beta$-alumina, different types of ion sites exist with different energies.\cite{bet69,fun84,fun06} In addition, the interionic Coulomb interactions cause a significant spread in the potential energies of the ions. In glasses, the disorder of the glass matrix leads to a broad distribution of ion site energies and barrier heights and thus to a broad distribution of jump rates.\cite{sva99,vog04a} This may explain why, even in single-modified glasses (i.e., with only one type of ion) with low number ion density and corresponding weak interionic Coulomb interactions, violations of TTS have not been observed.\cite{rol97,sid99a}

In contrast, crystals with low concentrations of point defects routinely show TTS violations. Examples are materials with intrinsic Frenkel or Schottky defects, such as alkali and silver halides.\cite{fun96a} In these materials, the interactions between the small number of defects are weak and the defects are partly bound to counter charges. Therefore, on short time scales the defects carry out localized movements close to the counter charges. These localized movements are not correlated to the long-range ion transport, and consequently the conductivity spectra do not obey TTS. 

Violations of TTS are also found in materials with more than one type of mobile ion. Examples are mixed-alkali glasses\cite{cra02}, as for instance 2 Ca(NO$_3$)$_2$ $\cdot$ 3 KNO$_3$ (CKN) melts.\cite{pim95,sin05} Below the glass transition temperature ($T_g=333$K) CKN is believed to be a pure K$^+$ ion conductor and it obeys TTS, but at higher temperatures Ca$^{2+}$ ions most likely also contribute significantly to the conductivity (above $375$K CKN again obeys TTS \cite{sin05}). Other examples are some polymer electrolytes above their glass-transition temperature where the polymer chains carry out segmental movements. Here different types of movements with different characteristic length scales contribute to the conductivity spectra, which generally results in TTS deviations.\cite{sid96}

\begin{figure}
\includegraphics[width=8cm]{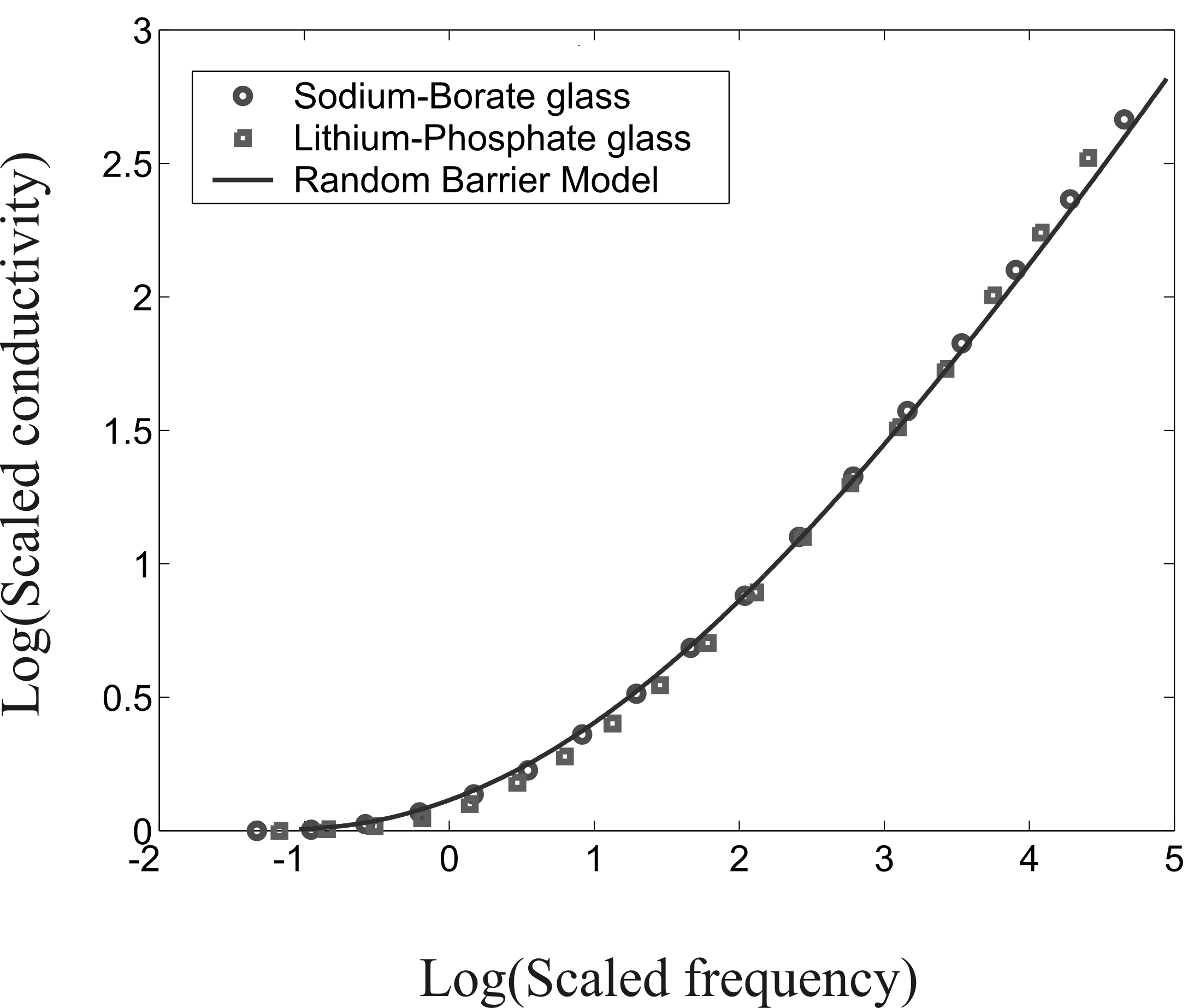}
\includegraphics[width=8cm]{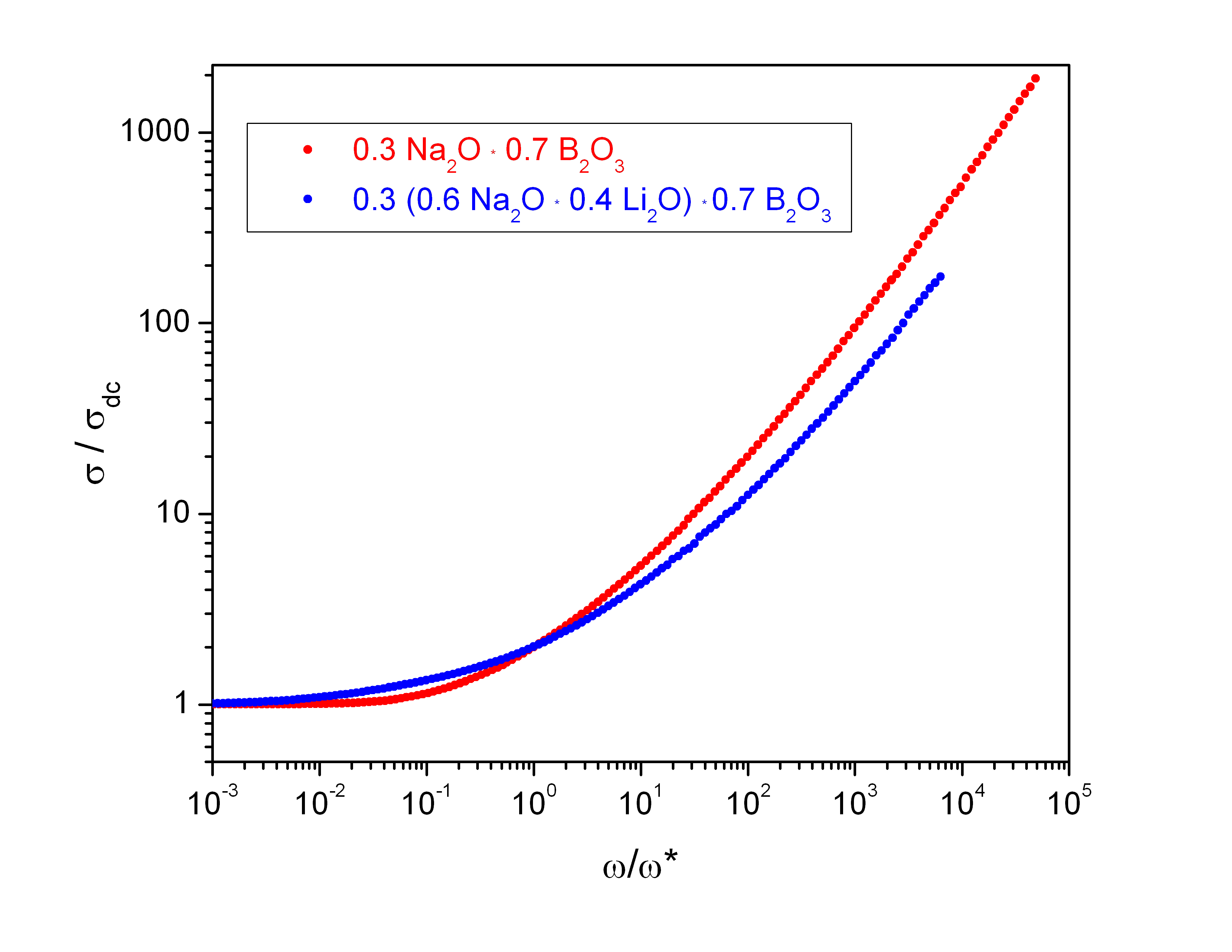}
\caption{Approximate ac universality and deviations from it. 
Figure 6(a) shows the RBM prediction for the scaled real part of the ac conductivity (full curve) compared to data for two typical ion conducting glasses, the $24^{o}$C data of Fig. 2(b) and the Sodium-Borate data of Fig. 6(b). The fit is good, but not perfect. 
Figure 6(b) shows that ac universality does not include mixed-alkali glasses (blue); the red curve for a Sodium-Borate glass represents approximate ac universality, compare Fig. 6(a).}
\end{figure}

The simplest model exhibiting the scaling properties Eqs. (\ref{eq:scaling}) and (\ref{eq:bnn}) is perhaps the random barrier model (RBM), see, e.g., Ref. \onlinecite{dyr00} for a review. In this model hopping of a single particle on a lattice with identical site energies is considered, where the energy barriers for jumps between neighboring sites are randomly drawn from a smooth probability distribution. The particles must overcome a critical ``percolation'' energy barrier $E_{\rm c}$ to exhibit long-range motion (a review of percolation theory with particular emphasis on ion diffusion was given by Bunde and Havlin \cite{bun96}). The time to overcome the percolation barrier, $t_{\rm c}\propto \exp(E_{\rm c}/k_{\rm B}T)$, determines the characteristic frequency marking onset of ac conduction: $\omega^\ast\sim t_{\rm c}^{-1}$.\cite{dyr00} The percolation energy barrier -- acting as a bottleneck -- also determines the dc conductivity temperature dependence. Thus percolation explains why a wide distribution of barriers nevertheless results in an Arrhenius dc conductivity (which is observed for most disordered ion-conducting solids). Incidentally, the BNN relation's rough proportionality, $\sigma(0)\propto\omega^\ast$, also follows from percolation determining the conduction properties. 

The scaling function of the RBM is {\it universal} in the ``extreme disorder limit'' where the jump rates vary over several decades; this limit is approached as temperature is lowered. Universality means that the ac response in scaled units becomes independent of both temperature and activation energy probability distribution. This was shown by extensive computer simulations involving barriers distributed according to a Gaussian, an exponential, an inverse power law, a Cauchy distribution, etc.\cite{dyr00,sch02} It was recently shown\cite{sch08} that if $\tilde\sigma\equiv \sigma(\omega)/\sigma_{dc}$ and $\tilde\omega$ is a suitably scaled frequency, except at low frequencies where the conductivity approaches the dc level, the universal RBM ac conductivity is to a good approximation given by the equation

\be
\label{eq:RBM}
\ln\tilde\sigma\,=\,
\left(\frac{i\tilde\omega}{\tilde\sigma}\right)^{2/3}\,.
\ee 
This expression implies an approximate power-law frequency dependence of the ac conductivity with an exponent that slowly converges to unity at very high frequencies  -- not simply an exponent of $2/3$ as one might naively guess. A very accurate representation of the RBM universal ac conductivity is given in Appendix B.

The RBM scaling function is usually close to, but rarely identical to those of experiments.\cite{dyr00,bar99, rol01a} As an example Fig. 6(a) shows the RBM universal ac conductivity (full curve) with the Lithium-Phosphate data of Fig. 2(b) at the lowest temperature ($24^{o}$C) and the Sodium-Borate data of Fig. 6(b), where both data sets were empirically scaled on the frequency axis. Figure 6(b) shows data also for a mixed-alkali glass; these data deviate significantly from the approximate ac universality represented by the red curve. It appears that, on the one hand, the RBM captures the essential features of the ion dynamics for single-ion conducting disordered solids, and that, on the other hand, deviations from the RBM universal ac conductivity may provide important information about specific features of the solid in question \cite{mur04}. Thus the RBM may be regarded as the ``ideal gas'' model for ac conduction in disordered solids.

In the RBM the dispersive transport properties are governed by strong disorder, forcing the ions to explore percolation paths for long-range motion.\cite{hun93,ish07} Macroscopic alternatives to the RBM (but with similar basic physics \cite{dyr00}), which apply if the sample has microstructure, have also been studied.\cite{dyr93,alm04} In most cases except that of nanocrystalline materials, however, the disorder is believed to be on the atomic scale.

Single-particle models like the RBM or its generalizations are simple and attractive for understanding the origin of TTS, but there are a number of challenges to this approach as well as open questions that one must keep in mind: 

\begin{itemize}
\item [{\it (i)}] For glasses the stoichiometry can be varied to a large extent. Related to changes in composition there are changes in activation energies, as for example a lowering of the activation energy with increasing mobile ion content or an increase of activation energy when one type of mobile ion is successively replaced by another type (the mixed-alkali effect, Sec. IV). These effects are not dealt with in the RBM unless the model is modified in an {\it ad hoc} manner to allow for significant changes in the barrier distribution and mismatch effects with respect to different ion types. The simplest models accounting for these effects are hopping systems with site exclusion,\cite{bar99,maa99} i.e., where there can be at most one ion at each site. Interestingly, such ``Fermionic'' hopping systems with site energy disorder often obey TTS.\cite{por00,pas06} Moreover, calculations for the corresponding single-particle systems yield scaling functions\cite{sch07} that are close to the RBM universal scaling function. A conclusive picture of the scaling properties of these type of models remains to be established, however.

\item[\it (ii)] Recent molecular dynamics simulations\cite{lam03,hab04,vog04b,lam05,mue07} and theories for the internal friction behavior in mixed-alkali glasses\cite{pei05} show that often only few of the potential ion sites are vacant (typically $1-3\%$). This is expected on general grounds, since during the cooling process a glass tends to a state of low free energy, thus with few defects.\cite{dyr03} It would be interesting to investigate whether hopping models with a low concentration of vacant sites generally obey TTS.

\item[\it (iii)] The Coulomb interactions between ions can be estimated from their mean distance $R\propto n^{-1/3}$, where  $n$ is the number density of 
ions. At room temperature, typical plasma parameters $e^2/(4\pi\epsilon_\infty R k_{\rm B}T)$ are in the range 30-80. In view of the confined geometry of the diffusion (percolation) paths\cite{mey04} the local interactions may be even stronger. Hence it is important to clarify whether hopping models with Coulomb interaction obey TTS and, if so, how the scaling function is affected by the interactions (for a general overview of Coulomb interactions effects on dispersive transport properties, see e.g. Ref. \onlinecite{die02}). Early studies of Coulomb interaction effects in hopping models with percolative disorder\cite{maa91} showed that Coulomb interactions give rise to a strong conductivity dispersion, but TTS was not observed. This might be due to the fact that in these early simulations temperature was not low enough. Another reason could be that, as in the RBM, a smooth and broad distribution of barrier or site energies is required for scaling. Indeed, simulation studies of many-particle hopping in the RBM with Coulomb interactions show agreement with the scaling behavior for low and moderate particle concentrations in the limit of low temperatures.\cite{rol01a} Overall, however, the problem is far from being settled; in particular if one takes into account that the fraction of empty sites should be small and that critical tests for other types of structural disorder such as site energy disorder, have not yet been performed. Due to the long-range nature of the Coulomb force, one could argue that its contribution to the energy landscape (sites and saddle points) provides an overall mean-field contribution. This hypothesis should be tested by further simulations.

\item[\it (iv)] Most studies of the RBM and other hopping models focused on site and/or barrier energies varying randomly without spatial correlation (a notable exception is the counterion model\cite{kno96,pen98}). If there are significant spatial correlations -- thus introducing a further length scale into the problem -- this may well lead to a breakdown of TTS.

\end{itemize}

\section{What causes the nearly constant loss?}

At high frequencies and/or low temperatures conductivity spectra approach a regime with nearly linear frequency dependence when plotted in the usual double-log plot: $\sigma'(\omega)\propto A \omega^n$ ($n\cong 1$). The proportionality constant $A$ is only weakly temperature dependent. This is referred to as the ``nearly constant loss'' (NCL) regime since it corresponds to an almost frequency-independent dielectric loss (Fig. 3). This behavior is ubiquitously observed in a wide variety of solids including glassy, crystalline, and molten ion conductors, independent of specific chemical and physical structures -- for an overview on experimental results, see e.g. Refs. \onlinecite{bur89,nga99}.

There are different possible origins of the NCL. One possibility is that NCL reflects the still not fully understood low-energy excitations present in all disordered materials. In the quantum-mechanical tunnelling model these excitations account for the anomalous low-temperature features of heat capacity and sound-wave absorption.\cite{phi87} At higher temperatures the low-energy excitations give rise to relaxations of the system over an energy barrier separating two different energy minima, described by the asymmetric double-well potential (ADWP) model.\cite{bal88} On a microscopic level this could correspond to cooperative ``jellyfish-type'' movements of groups of atomic species in the material.\cite{sid96,now94, lu94} If correct, this type of dynamic process should be a feature of all disordered materials, including materials without ions.

A more recent interpretation suggests that localized hopping movements of ions within fairly small clusters of sites contribute to the NCL in disordered ion conductors.\cite{rol01b,sid02} In this interpretation the NCL is merely the extension to higher frequencies of the dispersive conductivity. In fact, any hopping model with sufficient disorder gives rise to such a regime, since on short time scales hopping models always correspond to ADWP-type models. In the RBM, for instance, ion jumps over limited ranges lead to an NCL regime at high frequency, a region that extends to lower frequencies as temperature is lowered. There are two experimental observations favoring the second interpretation: (i) The magnitude of the NCL increases with increasing ion concentration\cite{sid02} (this also applies in the ADWP model if the defect centers are somehow associated with the mobile ions);  (ii) at large ion concentration and temperatures above 100 K, the scaling properties of the NCL contribution to the conductivity spectra are identical to the scaling properties found at lower frequencies where the dispersive conductivity passes over to the dc conductivity.\cite{rol01b,sid05} On the other hand, experiments carried out by one of the present authors suggest that the low-temperature ($T<$ 80 K) NCL in glasses with few ions is due to ADWP-type relaxation of the glass network.\cite{sid02} This indicates that both hopping movements of the ions and ADWP-type relaxations in the material contribute to the NCL.\cite{hsi96} Which of these dynamic processes dominates depends on composition and temperature.

Again, one may wonder whether it is permissible to neglect interactions, which can be modelled by dipolar forces as regards the short-time dynamics with only local movements of the mobile ions close to some counterions. Monte Carlo studies and analytical calculations have shown that the energetic disorder associated with the local electric field distributions of spatially randomly distributed dipoles gives rise to an NCL contribution at very low temperatures within an effective one-particle description, whereas at higher temperatures such behavior can occur due to many-particle effects.\cite{kno96,pen98,jai99,hoh02,sch04,maa06b,die08}

Finally, it has been suggested that the NCL is caused by vibrational movements of the mobile ions in strongly anharmonic potentials or from ion hopping in a slowly varying cage potential defined by neighboring mobile ions.\cite{leo01,nga02,riv02} These views focus on the very high-frequency NCL. Indeed, at frequencies in the THz range the ac conductivity joins into the vibrational absorption seen in far infrared spectroscopy associated with the quasi-vibrational motion of the mobile ion\cite{bur89}  Unfortunately, the connection between the NCL and the vibrational modes is poorly investigated: A data gap from the GHz to the THz regions exists because measurements of the dielectric loss are here particularly challenging.  More focused studies in this frequency window are needed to elucidate the connection between vibrational and librational (anharmonic) motion, as well as to better characterize the precise frequency dependence of the NCL conductivity, i.e., is it exactly linear ($n=1$), slightly sub-linear ($n<1,n\cong 1$) or slightly super-linear  ($n>1,n\cong 1$)?\cite{fun96,rit07,tiw07,lin08}

\section{What is the ion transport mechanism?}

\begin{figure}
\includegraphics[width=8cm]{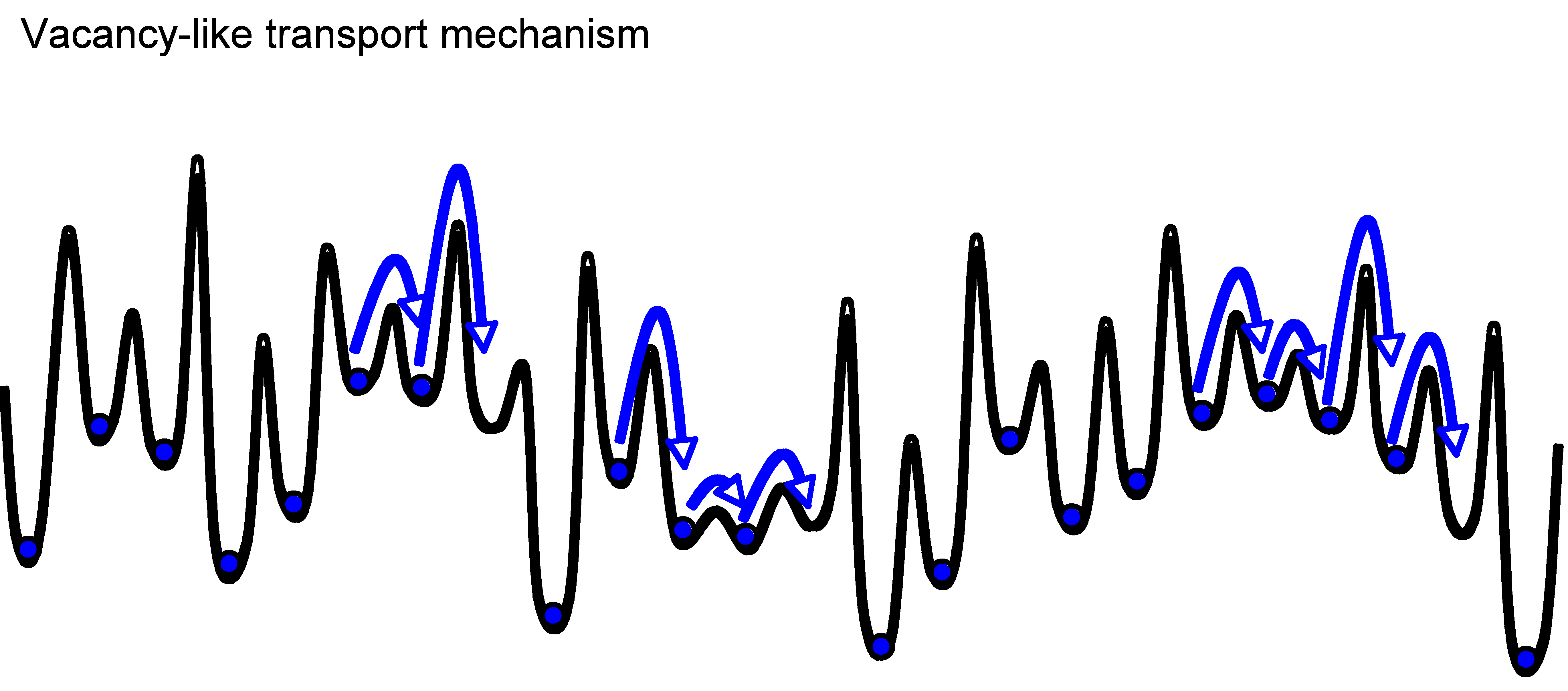}
\caption{Three vacancies jump to the left, each proceeding via a sequence of ion jumps to the right. First one ion jumps into the vacancy, then the next ion jumps into the new vacancy, etc.}
\end{figure}

We now turn briefly to the most fundamental question relating to ion conduction in disordered solids: What is the transport mechanism?\cite{owe63,tom77,ing87,ang90,ang92,nga96,mey99,kna00,fel02,kee02,ali06,dut07,hab07a,fun08a} As mentioned already, there is evidence that in many cases only few of the sites available for ions are actually vacant. Figure 7 illustrates the vacancy mechanism. Ion conduction in disordered solids does not proceed via the well-understood vacancy mechanism of ionic crystals with few vacancies.\cite{boh96} In crystals the vacancy concentration is strongly temperature dependent due to thermally activated defect formation; in glasses the concentration of empty sites is determined by the history of glass formation via the cooling rate, and the number of vacancies is frozen in at the glass transition \cite{dyr03}. Moreover, the vacancy concentration in glasses is believed to be significantly larger than in crystals. This makes the conduction mechanism in glasses much more complex, also because vacancy-vacancy interactions generally cannot be ignored. If such interactions are nevertheless not important, one may regard a vacancy as a charge carrier, e.g., in the RBM. In this case, the results of the RBM pertaining to the scaling features of conductivity spectra apply unaltered. 

Evidence for the significance of many-particle effects also comes from molecular dynamics simulations and from measurements of the Haven ratio. Simulations\cite{hab04,hab05,hab06,kun05} suggest that hopping occurs in a cooperative way, where one ion jump triggers jumps of other ions or hopping events occur collectively by involving several ions. A directional correlation of jumps of different ions is indicated by the measured Haven ratios that are generally smaller than unity,\cite{ter75} corresponding to positive cross-correlations in the current autocorrelation function. If heterogeneities in the host network confine the ion motion to channel-like structures, such correlations can be expected to be even more pronounced.\cite{mey04,bin05} Haven ratios smaller than unity have also been found in Monte Carlo simulations of models with Coulomb interactions.\cite{maa95}

There is a need for more systematic investigations of the role of many-particle effects. To uncover the ion transport mechanism(s) model predictions should be compared to other experimental observables than conductivity, such as NMR, spin-lattice relaxation, mechanical relaxation, tracer diffusion, quasi-elastic neutron scattering and multiple spin-echo experiments \cite{sva93,swe96,san08}. These methods probe different correlation functions, so checking model predictions against them obviously provides more severe tests than just, e.g., comparing predicted ac conductivity spectra to experiment. Such lines of inquiry, however, have so far only been undertaken in few instances; for example spin-lattice relaxation spectra were investigated in Refs. \onlinecite{mey93,boh07,maa95}, and multiple spin-echo experiments in Ref. \onlinecite{vog06}. -- The ion conduction mechanism clearly deserves a review on its own summarizing the latest developments. We have here mainly announced the problem and left out detailed considerations of, e.g., the role played by dynamic heterogeneities and possible dynamic channels for conduction pathways \cite{kun05,jun01}. Hopefully answering the other questions of the present review will provide valuable input into revealing the ion conduction mechanism.

\section{What is the role of dimensionality?}

The subdiffusive ion dynamics on transient time scales found ubiquitously in disordered materials is sometimes attributed to a fractal geometry of the conduction paths. A classic example of subdiffusion is particle dynamics occurring on a percolation cluster \cite{gef83,bun96}. In this case the cluster is fractal on length scales below a correlation length $\xi$, which diverges when approaching the percolation threshold. At criticality, dangling ends and loops occur on all lengths scales causing the mean-square displacement to increase as a power law with an exponent smaller than one. Close to the percolation threshold, $\xi$ is finite and subdiffusive behavior is observed only in an intermediate time regime, where the mean-square displacement is larger than microscopic length scales and smaller than $\xi^2$. For times where the mean-square displacement exceeds $\xi^2$, the diffusion eventually becomes normal. This examples suggests that the effective dimensionality of the conduction pathways may play an important role for the subdiffusive behavior, although for conduction pathways containing loops the fractal dimension of the pathway structure and the embedding Euclidian dimension of the material are generally not sufficient to determine the power law exponent in the subdiffusive time regime (see the discussion in chapter 3 of ref. \onlinecite{bun96}). Hence the question arises to what extent, if any, does the dimensionality of the conduction space influence the subdiffusive motion?

A possible scenario for qualitatively understanding the origin of the dimensionality dependence is the following. With decreasing dimensionality, the average distance between the highest barriers (percolation barriers) on the conduction pathways becomes larger. Between these percolation barriers, the ions perform back-and-forth motion. An increasing spatial extent of this back-and-forth motion leads to a larger dielectric relaxation strength, implying a more gradual transition from dc conductivity to dispersive conductivity in the conductivity spectrum.

Few studies of this question exist, but there is some evidence that dimensionality does influence the shape of the ac conductivity master curves (the scaling functions of Eq. (\ref{eq:scaling})). This can be seen in 2D crystals, like sodium $\beta-\mathrm{Al_2O_3}$, and in 1D crystals, like hollandite: The transition from dc conductivity to dispersive conductivity becomes more gradual with decreasing dimension.\cite{sid99b} This variation in the shape, as characterized by an effective exponent of Eq. (\ref{eq:RBM}), is shown in Fig. 8.  This sensitivity to dimensionality is also evident in the RBM for which the shape of the conductivity spectra are similarly altered by changing the dimensionality, i.e., in two dimensions the conductivity increases somewhat less steeply with frequency than in three dimensions.\cite{dyr96,note} 

\begin{figure}
\includegraphics[width=8cm]{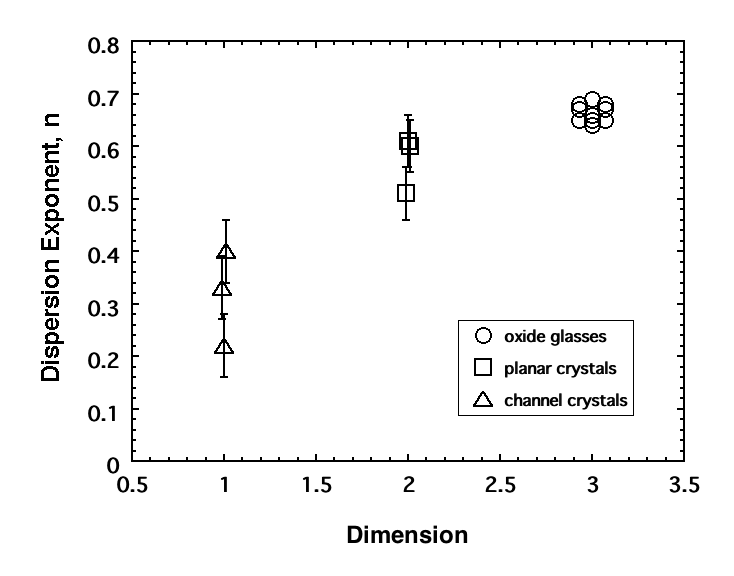}
\caption{Dimensionality dependence of the conductivity exponent $n$ (Eq. (\ref{jonscher})) for glasses, planar crystals and channel crystals.}
\end{figure}

The conducting pathways of the two above-mentioned crystals (sodium $\beta-\mathrm{Al_2O_3}$ and hollandite) have a well-defined dimensionality. Is there any other evidence for modifications of the correlated motion which might be connected to the dimensionality (localized) of the conduction space in an amorphous solid? While most disordered materials as mentioned show very similar shapes of the ac conductivity curves, some researchers\cite{sid00,rol00,ing03} have observed systematic changes that appear to arise from changes in the local environment of the ions. In studies of alkali-germanate glasses and alkali-borate glasses of varying ion content, for instance, subtle changes in the shape of the conductivity spectra were seen to correlate to known anomalies in the glass-transition temperature.\cite{rol01,mur04} The glass-transition temperature passes through a maximum as a result of how added modifier ions initially polymerize, but later depolymerize, the oxide network. Consequently, the average oxygen coordination of the ion's charge-compensating sites changes with ion concentration, resulting in modifications of the local ion environment which could mimic changes to the dimensionality of the conduction space.\cite{ani08} In a similar way, the mixing of two ion species (say Li and Na) modifies the local environment of the ion,\cite{ing03} and studies of ac conductivity of mixed-alkali glasses\cite{rol00} show a distinct change compared to that found in either single alkali end-member composition alone (compare Fig. 6(b)). Finally, in studies of metaphosphate glasses (whose oxide structure is highly polymeric) systematic changes in the correlated motion occurred in conjunction with variations in the cation size relative to the free volume \cite{sid03}. There it was posited that an effective local dimension of conduction space might rule the correlated motion. -- More studies along these lines are clearly warranted, but it appears that the effective dimensionality is an important parameter.

\section{Concluding remarks}

Science progresses by asking questions. It is our hope that this paper will stimulate to works focusing on basic understanding, eventually leading to a physical picture and quantitative model(s) of ion conduction in disordered solids that are as good as those of ionic and electronic conduction in crystals. It is a reasonable working hypothesis that ion conduction in disordered solids can be described in terms of a fairly simple generally applicable model, but only the future can tell whether this hope is realized.

\vspace{0.3cm}
{\bf Acknowledgements}

This paper was inspired by Round Table Discussions in 2007 at the {\it International Workshop on Ion Dynamics and Relaxation in Ion Conducting Disordered Solids} in Le Mans, France. The authors wish to thank Odile Bohnke, the organizer of this productive meeting.  Jeppe Dyre gratefully thanks ``Glass and Time,'' the Danish National Research Foundation's Centre for Viscous Liquid Dynamics, for support. Philipp Maass and Bernhard Roling gratefully acknowledge financial support by the HI-CONDELEC EU STREP project (NMP3-CT-2005-516975). Bernhard Roling would like to thank the German Science Foundation (DFG)   and the Alexander-von-Humboldt foundation for financial support of research projects. David Sidebottom would like to thank the U.S. Department of Energy, Division of Basic Energy Science (DE-FG03-98ER45696) for financial support.

\appendix\section{Relation between the long-time mean-square displacement and the low-frequency behavior of the ac conductivity}

The Kubo formula in dimension $d$ (where $s=i\omega+\epsilon$ is the ``Laplace frequency'' and $\epsilon >0$ is eventually taken to zero) reads

\begin{equation}
\sigma(\omega)=
\frac{1}{dVk_{\rm B}T}\int_0^\infty dt
\,\langle{\bf  I}(0)\cdot{\bf  I}(t)\rangle\,e^{-st}\,.
\label{eq:kubo1}
\end{equation}
Here $V$ is the sample volume and the total current ${\bf I}(t)$ is defined by summing over all $N$ ions:

\begin{equation}
{\bf  I}(t)=q\sum_{j=1}^{N}{\bf v}_j(t)\,.
\label{eq:current}
\end{equation}
Introducing the collective displacement over a time interval of length $t$,

\begin{equation}
\Delta{\bf {R}}(t)
=\sum_{j=1}^N\Delta{\bf  r}_j(t)
=\frac{1}{q}\int_0^tdt'{\bf I}(t')\,,
\label{eq:rvec}
\end{equation}
we have 

\begin{eqnarray}
q^2\langle\Delta{\bf{R}}^2(t)\rangle
&=&\int_0^tdt_1\int_0^tdt_2\,\langle{\bf I}(t_1)\cdot{\bf I}(t_2)\rangle
=2\int_0^td\tau\,(t-\tau)\langle{\bf I}(0)\cdot{\bf I}(\tau)\rangle\,,\nonumber\\
q^2\frac{d\langle\Delta{\bf {R}}^2(t)\rangle}{dt}
&=&2\int_0^td\tau\,\langle{\bf I}(0)\cdot{\bf I}(\tau)\rangle\,,\qquad
q^2\frac{d^2\langle\Delta{\bf{R}}^2(t)\rangle}{dt^2}
=2\langle{\bf I}(0)\cdot{\bf I}(t)\rangle\,.\nonumber
\end{eqnarray}
Accordingly, after a partial integration where the boundary term disappears because $\Delta{\bf{R}}^2(t)\sim t^2$ for $t\rightarrow 0$ (reflecting the short-time socalled ballistic motion), Eq.~(\ref{eq:kubo1}) takes the form 

\begin{equation}
\sigma(s)=\frac{q^2}{2dVk_{\rm B}T}\,
s\int_0^\infty dt\,\frac{d\langle\Delta{\bf R}^2(t)\rangle}{dt}
\,e^{-st}=C\,
s\int_0^\infty dt\,\dot f(t)
\,e^{-st}\,,
\label{eq:kubo2}
\end{equation}
where $C=nq^2/k_{\rm B}T$ and $f(t)=\langle\Delta{\bf R}^2(t)\rangle/2dN$.

We now make the ansatz

\begin{equation}
\sigma(s)\sim \sigma_{\rm dc}+As^\alpha\,,\hspace{2em}s\to0
\label{eq:aw}
\end{equation}
with $A$ real and $0<\alpha<1$. This is the well-known Jonscher ansatz \cite{jon77,jon83} analytically continued to complex frequencies, because for the real part of the frequency-dependent conductivity Eq. (\ref{eq:aw}) implies $\sigma'(\omega)\sim\sigma_{\rm dc}[1+(\omega/\omega^\ast)^\alpha]$ with 

\begin{equation}
\sigma_{\rm dc}\,(\omega^\ast)^{-\alpha}\,=\,
A\cos(\alpha\pi/2)\,.
\end{equation}
The analyticity requirement implies that at very low frequencies one must eventually have $\alpha=1$, but as an effective description of the regime of onset of ac conduction this ansatz may still be used. It follows that 

\begin{equation}
\frac{\sigma(s)-\sigma_{\rm dc}}{s}=
\int_0^\infty dt\,
\left[C\dot f(t)-\sigma_{\rm dc}\right] \,e^{-st}\sim As^{\alpha-1}\,,
\hspace{2em}s\to 0\,.
\end{equation}
Thus via a Tauberian theorem one concludes that

\begin{equation}
C\dot f(t)-\sigma_{\rm dc}
\sim \frac{A}{\Gamma(1-\alpha)}\,t^{-\alpha}\,,\hspace{2em}t\to\infty\,.
\label{eq:dr-asym}
\end{equation}
Since asymptotic expansions can be integrated term by term and $f(0)=0$, we obtain

\begin{equation}
Cf(t)\sim \sigma_{\rm dc}\,t+
\frac{A}{\Gamma(2-\alpha)}\,t^{1-\alpha}\,,\hspace{2em}t\to\infty
\end{equation}
or, if $D_\sigma\equiv\sigma_{\rm dc}/C=(k_{\rm B}T/nq^2)\sigma_{\rm dc}=\lim_{t\to\infty}\langle\Delta{\bf R}^2(t)\rangle/6Nt$ is a collective diffusion coefficient, 

\begin{equation}
\frac{\langle\Delta{\bf{R}}^2(t)\rangle}{2dN}\sim D_\sigma
\left[t+\frac{(\omega^\ast\,t)^{1-\alpha}}
{\omega^\ast\cos(\alpha\pi/2)\Gamma(2-\alpha)}\right]\,,\hspace{2em}t\to\infty\,.
\label{eq:rasym}
\end{equation}
If we introduce the time scale $t^\ast\equiv 1/\nu^\ast=2\pi/\omega^\ast$ corresponding to the above-defined crossover frequency $\omega^\ast$ where $\sigma'(\omega^\ast)=2\sigma_{\rm dc}$, and if we assume that the Jonscher ansatz is a good description near the crossover so that the asymptotic expression applies, Eq.~(\ref{eq:rasym}) implies

\begin{equation}
\sigma_{\rm dc}=\frac{nq^2}{2dk_{\rm B}T}
\frac{\langle\Delta{\bf {R}}^2(t^\ast)\rangle}{\gamma N}\,\nu^\ast
\end{equation}
with 

\begin{equation}
\gamma\,=\,
1+\frac{1}{(2\pi)^\alpha\cos(\alpha\pi/2)\Gamma(2-\alpha)}\,. 
\end{equation}
The factor $\gamma$ is roughly two for $\alpha\lesssim0.82$, but diverges for $\alpha\to1$. If we further replace the collective mean-square displacement $\langle\Delta{\bf {R}}^2(t^\ast)\rangle$ by the single-particle mean-square displacement $\langle\Delta {\bf r}^2(t^\ast)\rangle$, using the definition of the time-dependent Haven ratio

\begin{equation}
\frac{1}{H(t^\ast)}=1+\frac{1}{N}\frac{\sum_{j\ne k}
\langle\Delta{\bf r}_j(t^\ast)\cdot\Delta{\bf r}_k(t^\ast)\rangle}
{\langle\Delta {\bf r}^2(t^\ast)\rangle}
\end{equation}
that can be approximated by the Haven ratio $H$ in the dc-limit that is accessible via radioactive tracer experiments,\cite{hav65,ter75,isa99} i.e., $\langle\Delta{\bf{R}}^2(t^\ast)\rangle/N= \langle\Delta {\bf r}^2(t^\ast)\rangle/H(t^\ast)\simeq \langle\Delta {\bf r}^2(t^\ast)\rangle/H$, we arrive at Eq. (\ref{eq_sigdc1}) of the main text (where $d=3$):

\begin{equation}
\sigma_{\rm dc}=\frac{nq^2}{2dk_{\rm B}T}
\frac{\langle\Delta {\bf r}^2(t^\ast)\rangle}{\gamma H}\,\nu^\ast\,.
\end{equation}

\section{Accurate representation of the random barrier model universal ac conductivity}

The expression Eq. (\ref{eq:RBM}) gives a good overall fit to the universal ac conductivity of the random barrier model (RBM) arising in the extreme disorder limit, but in the range of frequencies where the conductivity approaches the dc level ($\omega<\omega^\ast$) there are significant deviations for both the real and imaginary parts of the conductivity. Thus for the imaginary part $\sigma''(\omega)$ Eq. (\ref{eq:RBM}) predicts that $\tilde\sigma''\propto\tilde\omega^{2/3}$ for $\tilde\omega\to 0$. This contradicts experiment, analyticity as well as RBM computer simulations\cite{sch08} that all imply $\tilde\sigma''\propto\tilde\omega$ for $\tilde\omega\to 0$. An accurate analytical representation of the RBM universal ac conductivity with the correct low-frequency behavior is given \cite{sch08} by the equation

\begin{equation}
\ln\tilde\sigma\,=\,
\frac{i\tilde\omega}{\tilde\sigma}\left(1+\frac{8}{3}\frac{i\tilde\omega}{\tilde\sigma}\right)^{-1/3}\,.
\end{equation}
In this expression frequency is scaled such that $\tilde\sigma=1+i\tilde\omega$ for $\tilde\omega\to 0$, i.e., the frequency scaling is different from that of Eq. (\ref{eq:RBM}). This equation is easily solved numerically for both real and imaginary parts as functions of frequency. Alternatively, numerical solutions to it -- as well as to Eq. (\ref{eq:RBM}) -- are available as ASCII files (see reference 22 of Ref. \onlinecite{sch08}).


\begin{thebibliography}{99}

\bibitem{kna00} Knauth P 2000 {\it J. Electroceram.} {\bf 5} 111 

\bibitem{dub03} Dubbe A 2003 {\it Sensors and Actuators B - Chemical} {\bf 88} 138

\bibitem{kha04} Kharton V V, Marques F M B and Atkinson A 2004 {\it Solid State Ionics} {\bf 174} 135 

\bibitem{kna04} Knauth P and Tuller H L 2004 {\it J. Am. Ceram. Soc.} {\bf 85} 1654 

\bibitem{vin06} Vinatier P and Hamon Y 2006 in {\it Charge Transport in Disordered Solids}, edited by S. Baranovski (Wiley, Chichester) p. 403

\bibitem{hui07} Hui SQ, Roller J, Yick S, Zhang X, Deces-Petit C, Xie Y S, Maric R and Ghosh D 2007 {\it J. Power Sources} {\bf 172} 493

\bibitem{nik07} Niklasson G A and Granqvist C G 2007 {\it J. Mater. Chem.} {\bf 17} 127

\bibitem{gro08} Grozema F C and Siebbeles L D A 2008 {\it Int. Rev. Phys. Chem.} {\bf 27} 87 

\bibitem{fun08} Funahashi M, Shimura H, Yoshio M and Kato T 2008 {\it Liquid crystalline functional assemblies and their supramolecular structures Book Series: Structure and bonding} {\bf 128} 151

\bibitem{bun98} Bunde A, Funke K and Ingram MD 1998 {\it Solid State Ionics} {\bf 105} 1

\bibitem{owe63} Owen A E 1963, in {\it Progress in Ceramic Science, Vol. 3}, edited by J. E. Burke (Macmillan, New York), p. 77

\bibitem{tom77} Tomozawa M 1977 in {\it Treatise on Materials Science, Vol. 12}, edited by M. Tomozawa (Academic, New York), p. 283

\bibitem{tul80} Tuller H L, Button D P and Uhlmann D R 1980 {\it J. Non-Cryst. Solids} {\bf 40} 93 

\bibitem{ing87} Ingram M D 1987 {\it Phys. Chem. Glasses} {\bf 28} 215

\bibitem{vin87} Vincent C A 1987 {\it Prog. Solid State Chem.} {\bf 17} 145

\bibitem{kre89} Kremer F, Dominguez L, Meyer W H and Wegner G 1989 {\it Polymer} {\bf 30} 2023 

\bibitem{ang90} Angell C A 1990 {\it Chem. Rev.} {\bf  90} 523 

\bibitem{mar91} Martin S W  1991 {\it J. Am. Ceramic Soc.} {\bf 74} 1767

\bibitem{ang92} Angell C A 1992 {\it Ann. Rev. Phys. Chem.} {\bf 43} 693

\bibitem{hei03} Heitjans P and Indris S 2003 {\it J. Phys.: Condens. Matter} {\bf 15}  R1257

\bibitem{mai95} Maier J 1995, {\it Prog. Solid State Chem.} {\bf 23} 171 

\bibitem{alm1983} Almond D P and West AR  1983 {\it Solid State lonics} {\bf 11} 57

\bibitem{dyr91} Dyre J C 1991 {\it J. Non-Cryst. Solids} {\bf 135}  219 

\bibitem{ell94} Elliott S R 1994 {\it J. Non-Cryst. Solids} {\bf 170} 97 

\bibitem{moy94} Moynihan C T 1994 {\it J. Non-Cryst. Solids} {\bf 172} 1395

\bibitem{nga00} Ngai K L and Rendell R W 2000 {\it Phys. Rev. B} {\bf 61} 9393

\bibitem{sid01} Sidebottom D L, Roling B and Funke K 2001 {\it Phys. Rev. B} {\bf 63} 024301

\bibitem{hod05} Hodge I M, Ngai K L and Moynihan C T 2005 {\it J. Non-Cryst. Solids} {\bf 351} 104

\bibitem{alm83a} Almond D P, Duncan G K and West A R 1983 {\it Solid State Ionics} {\bf 8} 159

\bibitem{alm83b} Almond D P and West A R 1983 {\it Solid State Ionics} {\bf 9/10} 277

\bibitem{hai94} Hairetdinov E F, Uvarov N F, Patel H K and Martin S W 1994 {\it Phys. Rev. B} {\bf 50} 13259

\bibitem{dyr88} Dyre J C 1988 {\it J. Appl. Phys.} {\bf 64} 2456

\bibitem{maa91} Maass P, Petersen J, Bunde A, Dieterich W and Roman H E 1991 {\it Phys. Rev. Lett.} {\bf 66}  52

\bibitem{fun93} Funke K 1993 {\it Prog. Solid St. Chem.} {\bf 22} 111

\bibitem{maa95} Maass P, Meyer M and Bunde A 1995 {\it Phys. Rev. B} {\bf 51} 8164

\bibitem{dyr00} Dyre J C and Schr{\o}der T B 2000 {\it Rev. Mod. Phys.} {\bf 72} 873

\bibitem{rol01} Roling B, Martiny C and Bruckner S 2001 {\it Phys. Rev. B} {\bf 63} 214203

\bibitem{mur04} Murugavel S and Roling B 2004 {\it J. Phys. Chem. B} {\bf 108} 2564

\bibitem{hav65} Haven Y and Verkerk B 1965 {\it Phys. Chem. Glasses} {\bf 6} 38

\bibitem{ter75} Terai R and Hayami R 1975 {\it J. Non-Cryst. Solids} {\bf 18} 217

\bibitem{isa99} Isard J O 1999 {\it J. Non-Cryst. Solids} {\bf 246} 16

\bibitem{sch94} Sch{\"u}tt H J 1994 {\it Solid State Ionics} {\bf 70/71} 505

\bibitem{tom98} Tomozawa M and Shin D-W 1998 {\it J. Non-Cryst. Solids} {\bf 241} 140

\bibitem{kle06} Klein R J, Zhang S H, Dou S, Jones B H, Colby R H and Runt J 2006 {\it J. Chem. Phys.} {\bf 124} 144903

\bibitem{mar08} Martin S W personal communication. 

\bibitem{riv04} Rivera A and Sanz J 2004 {\it Phys. Rev. B} {\bf 70} 094301

\bibitem{ber05} Berndt S, Jeffrey K R, K{\"u}chler R and B{\"o}hmer R 2005 {\it Solid State NMR} {\bf 27} 122

\bibitem{mus06} Mustarelli P, Tomasi C, Garcia M D P and Magistris A 2006 {\it Phys. Chem. Glasses -- Eur. J. Glass Sci. Technol. B} {\bf 47} 484 

\bibitem{boh03} Bohnke O, Badot J C and Emery J 2003 {\it J. Phys.: Condens. Matter} {\bf 15} 7571

\bibitem{boh07} B{\"o}hmer R, Jeffrey K R and Vogel M 2007 {\it Prog. Nucl. Magn. Res. Spect.} {\bf 50} 87

\bibitem{ing80} Ingram M D, Moynihan C T and Lesikar A V 1980 {\it J. Non-Cryst. Solids} {\bf 38-39} 371

\bibitem{rav77} Ravaine D and Souquet J L 1977 {\it Phys. Chem. Glasses} {\bf 18} 27

\bibitem{isa76} Isard J O 1976 {\it Phys. Chem. Glasses} {\bf 17} 1

\bibitem{hey77} Heyne L 1977 in {\it Solid Electrolytes}, edited by Geller S (Springer, Berlin) p.169 

\bibitem{rie91} Riess I 1991 {\it Solid State Ionics} {\bf 44} 199

\bibitem{rie97} Riess I 1997 in {\it CRC Handbook of Solid State Electrochemistry}, edited by Gellings P J and Bouwmeester H J M (CRC Press, New York), p. 223 

\bibitem{mac53} Macdonald J R 1953 {\it Phys. Rev.} {\bf 92} 4

\bibitem{fri54} Friauf R J 1954 {\it J. Chem. Phys.} {\bf 22} 1329

\bibitem {ils58} Ilschner B 1958 {\it J. Chem. Phys.} {\bf  28} 1109 

\bibitem{bea67} Beaumont J H and Jacobs P W M 1967 {\it J. Phys. Chem. Solids} {\bf 28} 657

\bibitem{baz04} Bazant MZ, Thornton K and Ajdari A 2004 {\it Phys. Rev. E} {\bf 70} 021506

\bibitem{sch92} Sch{\"u}tt H J and Gerdes E 1992 {\it J. Non-Cryst. Solids} {\bf 144} 14

\bibitem{pit03} Pitarch A, Bisquert J and Garcia-Belmonte G 2003 {\it J. Non-Cryst. Solids} {\bf 324} 196

\bibitem{hug68} Hughes K and Isard J O 1968 {\it Phys. Chem. Glasses} {\bf 9} 37

\bibitem{isa68} Isard J O 1968 {\it J. Non-Cryst. Solids} {\bf 1} 235

\bibitem{day76} Day D E  1976 {\it J. Non-Cryst. Solids} {\bf21}  343

\bibitem{ing94} Ingram M D 1994 {\it Glastech. Ber. Glass Sci. Technol.} {\bf 67} 151
  
\bibitem{vos04a} Voss S, Imre A W and Mehrer H 2004 {\it Phys Chem Chem Phys.} {\bf 6} 3669

\bibitem{vos04b} Voss S, Berkemeier F, Imre A W and Mehrer H 2004 {\it Z. Phys. Chem.} {\bf 218} 1353

\bibitem{maa92} Maass P, Bunde A and Ingram M D 1992 {\it Phys. Rev. Lett.} {\bf 68} 3064

\bibitem{bun94} Bunde A, Ingram M D and Maass P 1994 {\it J. Non-Cryst. Solids} {\bf 172-174} 1222

\bibitem{hun94} Hunt A 1994 {\it  J. Non-Cryst. Solids} {\bf 175} 129

\bibitem{gre95} Greaves G N and Ngai K L 1995 {\it Phys. Rev. B} {\bf 52}  6358

\bibitem{hab95} Habasaki J, Okada I and Hiwatari Y 1995 {\it J. Non-Cryst. Solids} {\bf 183} 12 

\bibitem{hun95} Hunt A 1995 {\it  J. Non-Cryst. Solids} {\bf 255} 47

\bibitem{hab96} Habasaki J, Okada I and Hiwatari Y 1996 {\it J. Non-Cryst. Solids} {\bf 208} 181 

\bibitem{swe98} Swenson J, Matic A, Brodin A, Borjesson L and Howells W S 1998 {\it Phys. Rev. B} {\bf 58} 11331

\bibitem{tom98b} Tomozawa M 1998 {\it Solid State Ionics} {\bf 105} 249

\bibitem{bar99} Baranovski S D and Cordes H 1999 {\it J. Chem. Phys.} {\bf 111} 7546

\bibitem{maa99} Maass P 1999 {\it J. Non-Cryst. Solids} {\bf 255} 35

\bibitem{sch99} Schulz B M, Dubiel M and Schulz M 1999 {\it  J. Non-Cryst. Solids} {\bf 241} 149

\bibitem{kir00} Kirchheim R 2000 {\it J. Non-Cryst. Solids}  {\bf 272}  85

\bibitem{swe01} Swenson J, Matic A, Karlsson C, B\"orjesson L, Meneghini C and Howells W S 2001 {\it Phys. Rev. B} {\bf 63} 132202
 
\bibitem{swe03} Swenson J and Adams S 2003 {\it Phys. Rev. Lett.} {\bf 90} 155507

\bibitem{bun04} Bunde A, Ingram M D and Russ S 2004 {\it Phys. Chem. Chem. Phys.} {\bf 6}  3663

\bibitem{hab04b} Habasaki J, Ngai K L and Hiwatari Y 2004 {\it J. Chem. Phys.} {\bf 121} 925
  
\bibitem{imr06} Imre A W, Divinski S V, Berkemeier F, and Mehrer H 2006 {\it J. Non-Cryst. Solids} {\bf 352} 783

\bibitem{hab07b} Habasaki J and Ngai K L 2007 {\it Phys. Chem. Chem. Phys.} {\bf 9} 4673

\bibitem{zie08} Zielniok D, Eckert H and Cramer C 2008 {\it Phys. Rev. Lett.} {\bf 100} 035901

\bibitem{gre91} Greaves G N, Gurman S J, Catlow C R A, Chadwick A V, Houde-Walter S, Henderson C M B and Dobson B R 1991 {\it Phil. Mag. A} {\bf 64} 1059

\bibitem{hou93} Houde-Walter S N, Inman J M, Dent A J and Greaves G N 1993 {\it J. Phys. Chem.} {\bf 97}  9330
  
\bibitem{gee97} Gee B, Janssen M and Eckert H 1997 {\it J. Non-Cryst. Solids} {\bf 215} 41

\bibitem{kam93} Kamitsos E I, Patsis A P and Chryssikos G D 1993 {\it J. Non-Cryst. Solids} {\bf 152} 246 

\bibitem{kam96} Kamitsos E I, Yiannopolous Y D, Jain H and Huang W C 1996 {\it Phys. Rev. B} {\bf 54} 9777

\bibitem{kam98} Kamitsos E I and Chryssikos G D 1998 {\it Solid State Ionics} {\bf 105} 75 
  
\bibitem{uch92} Uchino T, Sakka T, Ogata Y and Iwasaki M 1992 {\it J. Non-Cryst. Solids} {\bf 146} 26

\bibitem{uch99} Uchino T and Yoko T 1999 {\it J. Phys. Chem. B} {\bf 102}  8372
  
\bibitem{bal93} Balasubramanian S and Rao K J 1993 {\it J. Phys. Chem.} {\bf 97} 8835
  
\bibitem{lam03} Lammert H, Kunow M and Heuer A 2003 {\it Phys. Rev. Lett.} {\bf 90} 215901

\bibitem{lam05} Lammert H and Heuer A 2005 {\it Phys. Rev. B} {\bf 72} 214202 
 
\bibitem{pei05} Peibst R, Schott S and Maass P 2005 {\it Phys. Rev. Lett.} {\bf 95} 115901

\bibitem{maa06a} Maass P and Peibst R 2006 {\it J. Non-Cryst.Solids} {\bf 352} 5178

\bibitem{zda79} Zdaniewski W A, Rindone G E and Day D E 1979 {\it J. Mater. Sci.} {\bf 14}  763

\bibitem{dyr03} Dyre J C 2003 {\it J. Non-Cryst. Solids} {\bf 324} 192
 
\bibitem{hab04} Habasaki J and Hiwatari Y 2004 {\it Phys. Rev. B} {\bf 69} 144207

\bibitem{vog04b} Vogel M 2004 {\it Phys. Rev. B} {\bf 70} 094302

\bibitem{cha78} Chandrashekhar G V and Foster L M 1978 {\it Solid State Commun.} {\bf 27} 269

\bibitem{fos81} Foster L M, Anderson M P, Chandrashekhar G V, Burns G and Bradford R B 1981 {\it J. Chem. Phys.} {\bf 75} 2412 

\bibitem{bru83} Bruce J A and Ingram M D 1983 {\it Solid State Ionics} {\bf 9-10} 717 
  
\bibitem{mey96} Meyer M, Jaenisch V, Maass P and  Bunde A 1996 {\it Phys. Rev. Lett.} {\bf 76} 2338

\bibitem{mey98} Meyer M, Maass P and  Bunde A 1998 {\it  J. Chem. Phys.} {\bf  109} 2316

\bibitem{jon77} Jonscher A K 1977 {\it Nature} {\bf 267} 673
  
\bibitem{jon83} Jonscher A K 1996 {\it Universal Relaxation Law} (Chelsea Dielectrics Press, London)

\bibitem{fun07} Funke K, Singh P and Banhatti R D 2007 {\it Phys. Chem. Chem. Phys.} {\bf 9} 5582 

\bibitem{bow06} Bowen C R and Almond D P 2006 {\it Mater. Sci. and Technol.} {\bf 22} 719

\bibitem{mur07} Murugaraj R 2007 {\it J. Mater. Sci.} {\bf 42} 10065
 
\bibitem{pap07} Papathanassiou A N, Sakellis I and Grammatikakis J 2007 {\it Appl. Phys. Lett.} {\bf 91} 122911

\bibitem{bar66} Barton J L 1966 Verres Refr. {\bf 20} 328
  
\bibitem{nak72} Nakajima T 1972 in {\it Annual Report, Conference on Electric Insulation and Dielectric Phenomena} (National Academy of Sciences, Washington DC) p. 168
  
\bibitem{nam75} Namikawa H 1975 {\it J. Non-Cryst. Solids} {\bf 18} 173

\bibitem{dyr86} Dyre J C 1986 {\it J. Non-Cryst. Solids} {\bf 88} 271

\bibitem{sch00} Schr{\o}der T B and Dyre J C 2000 {\it Phys. Rev. Lett.} {\bf  84} 310

\bibitem{die02} Dieterich W and Maass P 2002 {\it Chem. Phys.} {\bf 28} 439

\bibitem{bet69} Bettman M and Peters C R 1969 {\it J. Phys. Chem.} {\bf 73} 1774

\bibitem{fun84} Funke K and Schneider H J 1984 {\it Solid State Ionics} {\bf 13} 335

\bibitem{fun06} Funke K, Banhatti R D, Wilmer D, Dinnebier R, Fitch A and Jansen M 2006 {\it J. Phys. Chem. A} {\bf 110} 3010 

\bibitem{sva99} Svare I 1999 {\it Solid State Ionics} {\bf 125} 47

\bibitem{vog04a} Vogel M, Brinkmann C, Eckert H and Heuer A 2004 {\it Phys. Rev. B} {\bf 69} 094302

\bibitem{rol97} Roling B, Happe A, Funke K and Ingram M D 1997 {\it Phys. Rev. Lett.} {\bf 78} 2160 (in this work the BNN relation was not 
used to scale the spectra, but the authors used $\omega^{\ast}\propto \sigma_{\mathrm dc}T/n$ that corresponds to the BNN crossover frequency if 
$\Delta\epsilon$ follows a Curie law, $\Delta\epsilon\propto n/T$).

\bibitem{sid99a} Sidebottom D L 1999 {\it Phys. Rev. Lett.} {\bf 82} 3653

\bibitem{fun96a} Funke K, Wilmer D, Lauxtermann T, Holzgreve R and Bennington S M 1996 {\it Solid State Ionics} {\bf 86-88} 141

\bibitem{cra02} Cramer C, Bruckner S, Gao Y  and Funke K 2002 {\it Phys. Chem. Chem. Phys.} {\bf 4} 3214

\bibitem{pim95} Pimenov A, Lunkenheimer P, Rall H, Kohlhaas R, Loidl A and B{\"o}hmer R 1995 {\it Phys. Rev. E} {\bf 54} 676

\bibitem{sin05} Singh P, Banhatti R D and Funke K 2005 {\it Phys. Chem. Glasses} {\bf 46} 241

\bibitem{sid96} Sidebottom D L, Green P F and Brow R K 1996 {\it J. Non-Cryst. Solids} {\bf 203} 300

\bibitem{bun96} Bunde A and Havlin S 1996 {\it Fractals and Disordered Systems} (Springer, Berlin)

\bibitem{sch02} Schr{\o}der T B and Dyre J C 2002 {\it Phys. Chem. Chem. Phys.} {\bf 4} 3173

\bibitem{sch08} Schr{\o}der T B and Dyre J C 2008 {\it Phys. Rev. Lett.} {\bf 101}, 025901

\bibitem{rol01a} Roling B 2001 {\it Phys. Chem. Chem. Phys.} {\bf 3}  5093

\bibitem{hun93} Hunt A 1993 {\it J. Non-Cryst. Solids} {\bf 160} 183

\bibitem{ish07} Ishii T 2007  {\it J. Phys. Soc. Japan} {\bf 76} 064603

\bibitem{dyr93} Dyre J C 1993 {\it Phys. Rev. B} {\bf 48} 12511

\bibitem{alm04} Almond D P and Bowen C R 2004 {\it Phys  Rev. Lett.} {\bf 92} 157601

\bibitem{por00} Porto M, Maass P, Meyer M , Bunde A and Dieterich W 2000 {\it Phys. Rev. B} {\bf 61} 6057

\bibitem{pas06} Pasveer W F, Bobbert P A and Michels M A J 2006 {\it Phys. Rev. B} {\bf 74} 165209

\bibitem{sch07} Schr{\o}der T B 2008 {\it Europhys. Lett.} {\bf 81} 30002

\bibitem{mue07} M\"uller C, Zienicke E, Adams S, Habasaki J and Maass P 2007 {\it Phys. Rev. B} {\bf 75} 014203

\bibitem{mey04} Meyer A, Horbach J, Kob W, Kargl F and Schober H 2004 {\it Phys. Rev. Lett.} {\bf 93} 027801

\bibitem{kno96} Kn{\"o}dler D, Pendzig P and Dieterich W 1996 {\it Solid State Ionics} {\bf 86-88} 29

\bibitem{pen98} Pendzig P  and Dieterich W 1998 {\it Solid State Ionics} {\bf 105} 209

\bibitem{bur89} Burns A, Chryssikos G D, Tombari E, Cole R H and Risen W M 1989 {\it Phys. Chem. Glasses} {\bf 30} 264

\bibitem{nga99} Ngai K L 1999 {\it J.Chem. Phys.} {\bf 110} 10576

\bibitem{phi87} Phillips W A 1987 {\it Rep. Progr. Phys.} {\bf 50} 1657 

\bibitem{bal88} Balzer-J{\"o}llenbeck G, Kanert O, Steinert J and Jain H 1988 {\it Solid State Comm.} {\bf 65} 303

\bibitem{now94} Nowick A S, Lim B S and Vaysleyb A V 1994 {\it J. Non-Cryst. Solids} {\bf 172} 1243

\bibitem{lu94}  Lu X and Jain H 1994 {\it J. Phys. Chem. Solids} {\bf 55} 1433

\bibitem{rol01b} Roling B, Martiny C and Murugavel S 2001 {\it Phys. Rev. Lett.} {\bf 87} 085901

\bibitem{sid02} Sidebottom D L and Marray-Krezan C M 2002 {\it Phys. Rev. Lett.} {\bf 89} 195901

\bibitem{sid05} Sidebottom D L 2005 {\it Phys. Rev. B} {\bf 71} 134206

\bibitem{hsi96} Hsieh C H and Jain H 1996 {\it J. Non-Cryst. Solids} {\bf 203} 293

\bibitem{jai99} Jain H 1999 {\it Met. Mater. Processes} {\bf 11} 317

\bibitem{hoh02} H\"ohr T, Pendzig P, Dieterich W and Maass P 2002 {\it Phys. Chem. Chem. Phys.} {\bf 4} 3168

\bibitem{sch04} Schulz M, Dieterich W and Maass P 2004 {\it Z. Phys. Chem.} {\bf 218} 1375

\bibitem{maa06b} Maass P, Dieterich W and Scheffler F 2006, in {\it Flow Dynamics: The Second International Conference on Flow Dynamics}, AIP Conference Proceedings, edited by M. Tokuyama and S. Maruyama (Melville, New York) {\bf 832} 492.

\bibitem{die08} Dieterich W, Maass P and Schulz M 2008 {\it Eur. Phys. J.: Special Topics} {\bf 161} 79

\bibitem{leo01} Leon C, Rivera A, Varez A, Sanz J, Santamaria J and Ngai KL 2001 {\it Phys. Rev. Lett.} {\bf 86} 1279

\bibitem{nga02} Ngai K L and Leon C 2002 {\it Phys. Rev. B} {\bf 66} 064308

\bibitem{riv02} Rivera A, Leon C, Varsamis C P E, Chryssikos G D, Ngai K L, Roland C M and Buckley L J 2002 {\it Phys. Rev. Lett.} {\bf 88} 125902

\bibitem{fun96} Funke K, Cramer C, Roling B, Saatkamp T, Wilmer D and Ingram M D 1996 {\it Solid State Ionics} {\bf 85} 293

\bibitem{rit07} Ritus A I 2007 {\it J. Phys.: Condens. Matter} {\bf 19} 086222

\bibitem{tiw07} Tiwari J P and Shahi K 2007 {\it Phil. Mag. } {\bf 87} 4475

\bibitem{lin08} Linares A, Canovas M J and Ezquerra T A 2008  {\it J. Chem. Phys.} {\bf 128} 244908

\bibitem{nga96} Ngai K L 1996 {\it J. Non-Cryst. Solids} {\bf 203} 232

\bibitem{mey99} Meyer W H 1999 {\it Adv. Materials} {\bf 10} 439

\bibitem{fel02} Feldman Y, Puzenko A and Ryabov Y 2002 {\it Chem. Phys.} {\bf 284} 139

\bibitem{kee02} Keen D A 2002 {\it J. Phys.: Condens. Matter } {\bf 14} R819

\bibitem{ali06} Alig I, Dudkin S A, Jenninger W and Marzantowicz M 2006 {\it Polymer} {\bf 47} 1722

\bibitem{dut07} Dutta A and Ghosh A 2007 {\it J. Chem. Phys.} {\bf 127} 144504

\bibitem{hab07a} Habasaki J 2007 {\it J. Non-Cryst. Solids} {\bf 353} 3956

\bibitem{fun08a} Funke K and Banhatti R D 2008 {\it Solid State Sciences} {\bf 10} 790

\bibitem{boh96} Bohnke O, Bohnke C and Fourquet J L 1996 {\it Solid State Ionics} {\bf 91} 21

\bibitem{hab05} Habasaki J, Ngai K L and Hiwatari Y 2005 {\it J. Chem. Phys.} {\bf 122} 054507

\bibitem{hab06} Habasaki J and Ngai K L 2006 {\it J. Non-Cryst. Solids} {\bf 352} 5170

\bibitem{kun05} Kunow M and Heuer A 2005 {\it Phys. Chem. Chem. Phys.} {\bf 7} 2131

\bibitem{bin05} Binder K, Horbach J, Winkler A and Kob W 2005 {\it Ceram. Int.} {\bf 31} 713

\bibitem{sva93} Svare I, Borsa F, Torgeson D R and Martin S W 1993 {\it Phys. Rev. B} {\bf 48} 9336

\bibitem{swe96} J Swenson and Borjesson L 1996 {\it Phys. Rev. Lett.} {\bf 77} 3569

\bibitem{san08} Sangoro J R, Serghei A, Naumov S, Galvosas P, Karger J, Wespe C, Bordusa F, and Kremer F 2008 {\it Phys. Rev. E} {\bf 77} 051202

\bibitem{mey93} Meyer M, Maass P and Bunde A 1993 {\it Phys. Rev. Lett.} {\bf 71} 573

\bibitem{vog06} Vogel M, Brinkmann C, Eckert H and Heuer A 2006 {\it J. Non-Cryst. Solids} {\bf 352} 5156

\bibitem{jun01} Jund P, Kob W and Jullien R 2001 {\it Phys. Rev. B} {\bf  64} 134303

\bibitem{gef83} Gefen Y, Aharony A and Alexander S 1983 {\it Phys. Rev. Lett.} {\bf 50} 77

\bibitem{sid99b} Sidebottom D L 1999 {\it Phys. Rev. Lett.} {\bf 83} 983

\bibitem{dyr96} Dyre J C and Schr{\o}der T B 1996 {\it Phys. Rev. B} {\bf 54} 14884

\bibitem{note} In Ref. \onlinecite{dyr96} ac conduction in one dimension was also simulated. This was done by introducing an artificial activation energy cut-off, which is needed to get sensible results (in two and three dimensions the cut-off is provided automatically by the percolation phenomenon). In this artificial model the ac conductivity is more like it is in three dimensions, than like two dimensions. One dimension is often strange and unphysical, however, so we do not believe that this observation invalidate the general argument presented.

\bibitem{rol00} Roling B and Martiny C 2000 {\it Phys. Rev. Lett.} {\bf 85} 1274

\bibitem{ing03} Ingram M D and Roling B 2003 {\it J. Phys.: Condens. Matter} {\bf 15} S1595

\bibitem{sid00} Sidebottom D L 2000  {\it Phys. Rev. B} {\bf 61} 14507

\bibitem{ani08} Aniya M 2008 {\it J. Non-Cryst. Solids} {\bf 354} 365

\bibitem{sid03} Sidebottom D L 2003 {\it J. Phys.: Condens. Matter} {\bf 15} S1585 

\end{thebibliography}
\end{document}